\documentclass[]{llncs}
\def\VeryFinalVersion{}

\def\AuthorVersion{}

\ifdefined\AuthorVersion
\else
\makeatletter
\AtBeginDocument{%
  \@ifpackageloaded{hyperref}
  {\def\@doi#1{\href{https://doi.org/#1}
      {\ttfamily https://doi.org/#1}\egroup}}
  {\def\@doi#1{\ttfamily https://doi.org/#1\egroup}}
  \def\doi{\bgroup\catcode`\_=12\relax\@doi}}
\makeatother
\fi

\usepackage[utf8]{inputenc}
\usepackage[dvipsnames,svgnames,table]{xcolor}  %
\usepackage{xspace}
\usepackage[%
		pdfauthor={Étienne André, Michał Knapik, Didier Lime, Wojciech Penczek and Laure Petrucci},%
		pdftitle={Parametric Verification: An Introduction},
	breaklinks  = true,
            colorlinks  = true,
            linkcolor   = black,
            urlcolor    = blue!50!green,
            citecolor   = Blue!50!blue,
            anchorcolor = Green]{hyperref}
\usepackage[super]{nth}  %

\usepackage{tikz}
\usepackage{subcaption}

\usepackage{pifont}
\usepackage{mathtools}
\usepackage[capitalise,english,nameinlink]{cleveref}
\usetikzlibrary{calc}
\usetikzlibrary{arrows,shapes,automata,backgrounds,petri,intersections}
\usetikzlibrary{decorations}
\usetikzlibrary{decorations.pathmorphing}
\usetikzlibrary{shadows}
\usetikzlibrary{shapes.symbols}
\usetikzlibrary{shapes.callouts}
\usetikzlibrary{patterns}
\usetikzlibrary{positioning}
\usetikzlibrary{backgrounds}
\tikzstyle{place}=[circle,thick,draw=blue!75,fill=blue!20,minimum size=6mm]
\tikzstyle{contour place}=[place,draw=red!100]
\tikzstyle{red place}=[place,draw=red!75,fill=red!20]
\tikzstyle{gray place}=[place,draw=black!100,fill=black!30]
\tikzstyle{transition}=[ rectangle,thick, fill=black, minimum width=8mm, inner ysep=2pt]	    
\tikzstyle{red transition}=[ rectangle,thick, fill=red!75, minimum width=8mm, inner ysep=2pt]	    
\tikzstyle{gray transition}=[ rectangle,thick, draw=black!100,fill=black!30, minimum width=8mm, inner ysep=2pt]	    
\tikzstyle{blue transition}=[ rectangle,thick, fill=blue!75, minimum width=8mm, inner ysep=2pt]	    
\tikzstyle{every label}=[black]
\tikzstyle{every node}=[initial text=]
\tikzstyle{location}=[rectangle, rounded corners, minimum size=12pt, draw=black, fill=blue!10, inner sep=2pt]
\tikzstyle{location10}=[location, minimum size=10pt]
\tikzstyle{invariant}=[draw=black, dotted, inner sep=1pt] %

\tikzstyle{goodtile} = [draw=green!50!black, fill=green]
\tikzstyle{badtile} = [draw=red!50!black, fill=red]

\tikzstyle{invisible}=[draw=none]
\tikzstyle{final}=[double]
\tikzstyle{urgent}=[fill=yellow!50]
\tikzstyle{bad}=[fill=red!50]

\usepackage{amsmath} %
\usepackage{amssymb} %

\newtheorem{notation}{Notation}

\ifdefined \VersionWithComments
  \usepackage{marginnote}
  \newcommand{\marginX}{\marginnote{\huge{\quad\textbf{!}\quad}}}
  \newcommand{\ea}[1]{\textcolor{purple}{\marginX{}[\textbf{Étienne}: #1]}}
  \newcommand{\mk}[1]{\textcolor{magenta}{\marginX{}[\textbf{Michał}: #1]}}
  
  \newcommand{\lp}[1]{\textcolor{green!70!black}{\marginX{}[\textbf{Laure}: #1 ]}}
  \newcommand{\wop}[1]{\textcolor{cyan!75!black}{\marginX{}[\textbf{Wojciech}: #1
  ]}}
  \newcommand{\todo}[1]{\textcolor{red}{\marginX{}TODO: #1}}
\else
  \newcommand{\ceb}[1]{}
  \newcommand{\ea}[1]{}
  \newcommand{\ja}[1]{}
  \newcommand{\mk}[1]{}
  \newcommand{\ms}[1]{}
  \newcommand{\lp}[1]{}
  \newcommand{\wop}[1]{}
  \newcommand{\todo}[1]{}
\fi

\newcommand{\eg}{\emph{e.g.}\xspace}
\newcommand{\ie}{\emph{i.e.}\xspace}
\newcommand{\viz}{viz.\xspace}
\newcommand{\wrt}{\emph{w.r.t.}\xspace}

\definecolor{cv1}{rgb}{1, 0, 0}
\definecolor{cv2}{rgb}{0, 1, 0}
\definecolor{cv3}{rgb}{0, 0, 1}
\definecolor{cv4}{rgb}{1, 1, 0}
\definecolor{cv5}{rgb}{1, 0, 1}
\definecolor{cv6}{rgb}{0, 1, 1}
\definecolor{cv7}{rgb}{0.8, 0.6, 0.4}
\definecolor{cv8}{rgb}{0.5, 0.5, 1}
\definecolor{cv9}{rgb}{0.55, 0.75, 0.35}
\definecolor{cv10}{rgb}{1, 0.6, 0.1}
\definecolor{cv11}{rgb}{0.6, 0.7, 0.8}
\definecolor{cv12}{rgb}{0.2, 0.5, 0.9}
\definecolor{cv13}{rgb}{0.5, 0.9, 0.2}
\definecolor{cv14}{rgb}{1, 0.3, 0.5}
\definecolor{cv15}{rgb}{0.7, 0.7, 0.7}

\definecolor{couleuract}{rgb}{0.50, 0.70, 0.30}
\definecolor{couleuractset}{rgb}{1, 0.20, 0.10}
\definecolor{couleurclock}{rgb}{0.4, 0.4, 1}
\definecolor{couleurcode}{rgb}{0.5, 0.2, 0.9} %
\definecolor{couleurdisc}{rgb}{1, 0, 1}
\definecolor{couleurloc}{rgb}{0.4, 0.4, 0.65}
\definecolor{couleurparam}{rgb}{1, 0.6, 0.0}
\definecolor{couleurpreuve}{rgb}{1, 0, 0}
\definecolor{couleurprob}{rgb}{1, 0, 0}
\definecolor{couleurref}{rgb}{0.2, 0.3, 1}

\definecolor{inputcolor}{rgb}{0.0, 0.6, 0.0}
\definecolor{outputcolor}{rgb}{0.7, 0.1, 0.7}

\definecolor{couleurbad}{rgb}{0.8, 0, 0}
\definecolor{couleurbadalt}{rgb}{0.6, 0.4, 0.4}
\definecolor{couleurgood}{rgb}{0, 0.7, 0}

\definecolor{couleurfondtrestresclair}{rgb}{0.92, 0.92, 1}

\newcommand{\styleact}[1]{\ensuremath{\textcolor{coloract}{\mathit{#1}}}}
\newcommand{\styleclock}[1]{\ensuremath{\textcolor{colorclock}{#1}}}

\newcommand{\styleloc}[1]{\textcolor{colorloc}{\ensuremath{\mathrm{#1}}}}
\newcommand{\styleparam}[1]{\ensuremath{\textcolor{colorparam}{#1}}}
\newcommand{\styleprop}[1]{\textcolor{colorloc}{\ensuremath{\mathbf{#1}}}}

\definecolor{coloract}{rgb}{0.50, 0.70, 0.30}
\definecolor{colorclock}{rgb}{0.4, 0.4, 1}
\definecolor{colorconst}{rgb}{0.50, 0.20, 0.00}
\definecolor{colordisc}{rgb}{1, 0, 1}
\definecolor{colorloc}{rgb}{0.4, 0.4, 0.65}
\definecolor{colorparam}{rgb}{1, 0.6, 0.0}

\tikzstyle{pzgState} = [
rectangle split,
rectangle split horizontal,
rectangle split parts=2,
draw,
rectangle split part align={center,left},
node distance=3.5 cm,
rounded corners,
minimum height=0.5 cm,
minimum width=1.5 cm,
thick
]

\newcommand{\init}{_0}

\newcommand{\A}{\ensuremath{\mathcal{A}}}

\newcommand{\Actions}{\Sigma}
\newcommand{\action}{\ensuremath{\sigma}}

\newcommand{\AP}{\ensuremath{\mathit{AP}}}
\newcommand{\atomicprop}{\ensuremath{\textit{ap}}}

\newcommand{\BTrue}{\text{true}}

\newcommand{\bounds}{\mathit{bounds}}
\newcommand{\boundinf}{\mathrm{inf}}
\newcommand{\boundsup}{\mathrm{sup}}
\newcommand{\C}{C}

\newcommand{\Clock}{X} %
\newcommand{\clocki}[1]{\styleclock{\clock_{#1}}}
\newcommand{\ClockCard}{H} %
\newcommand{\clock}{x} %
\newcommand{\clockval}{w} %
\newcommand{\ClocksZero}{\vec{0}}
\newcommand{\compOp}{\bowtie}

\newcommand{\edge}{e}
\newcommand{\Edges}{E}
\newcommand{\longuefleche}[1]{\stackrel{#1}{\mapsto}}
\newcommand{\longueflecheRel}{{\mapsto}}
\newcommand{\fleche}[1]{\stackrel{#1}{\rightarrow}}
\newcommand{\flecheRel}{{\rightarrow}}
\newcommand{\Fleche}[1]{\stackrel{#1}{\Rightarrow}}
\newcommand{\formule}{\ensuremath{\varphi}} %
\newcommand{\grandn}{\ensuremath{\mathbb N}}
\newcommand{\grandq}{\ensuremath{\mathbb Q}}
\newcommand{\grandqplus}{\ensuremath{\grandq_{+}}} %
\newcommand{\grandr}{\ensuremath{\mathbb R}}
\newcommand{\grandrplus}{\ensuremath{\grandr_{+}}} %
\newcommand{\grandz}{\ensuremath{\mathbb Z}}
\newcommand{\guard}{g}
\newcommand{\invariant}{I}
\newcommand{\K}{K}

\newcommand{\loc}{l} %

\newcommand{\locinit}{\loc\init}
\newcommand{\LocFinal}{F}
\newcommand{\Loc}{L} %

\newcommand{\lterm}{\mathit{lt}}

\newcommand{\Param}{P} %
\newcommand{\param}{p} %
\newcommand{\parami}[1]{\styleparam{\param_{#1}}}
\newcommand{\ParamCard}{M} %

\newcommand{\pval}{v} %

\newcommand{\plterm}{\mathit{plt}}
\newcommand{\PZG}{\ensuremath{\mathcal{PZG}}} %
\newcommand{\varproblem}{\ensuremath{\phi}}
\newcommand{\Problem}{\ensuremath{\mathcal{P}}}
\newcommand{\resets}{R}

\newcommand{\sinit}{s\init} %

\renewcommand{\S}{S}
\newcommand{\somelocs}{T} %
\newcommand{\state}{\ensuremath{s}} %
\newcommand{\States}{S} %
\newcommand{\symbstate}{\ensuremath{\mathbf{s}}} %
\newcommand{\SymbState}{\ensuremath{\mathbf{S}}} %
\newcommand{\symbstateinit}{\symbstate\init} %
\newcommand{\symbtrans}{{\Rightarrow}} %
\newcommand{\Succ}{\mathsf{Succ}}
\newcommand{\Time}{\mathsf{time}}
\newcommand{\timelapse}[1]{#1^\nearrow}

\newcommand{\UL}{\ensuremath{\mathsf{UL}}}

\newcommand{\mitlzeroinf}{{\ensuremath{\mathsf{MITL}_{0,\infty}}}}
\newcommand{\pmitlzeroinf}{{\ensuremath{\mathsf{PMITL}_{0,\infty}}}}

\newcommand{\bounded}[2]{#1_{|#2}}
\newcommand{\boundinfin}[2]{\boundinf(#1,#2)}
\newcommand{\boundsupin}[2]{\boundsup(#1,#2)}

\newcommand{\reset}[2]{\ensuremath{[#1]_{#2}}}
\newcommand{\valuate}[2]{\ensuremath{#2(#1)}}
\newcommand{\wv}[2]{#1|#2} %

\newcommand{\styleCTL}[1]{\ensuremath{\textsf{#1}}}
\newcommand{\EC}{\styleCTL{EC}}
\newcommand{\ED}{\styleCTL{ED}}
\newcommand{\EF}{\styleCTL{EF}}
\newcommand{\EG}{\styleCTL{EG}}
\newcommand{\AF}{\styleCTL{AF}}
\newcommand{\AG}{\styleCTL{AG}}
\newcommand{\CTLA}{\styleCTL{A}}
\newcommand{\CTLE}{\styleCTL{E}}
\newcommand{\CTLF}{\styleCTL{F}}
\newcommand{\CTLG}{\styleCTL{G}}
\newcommand{\CTLX}{\styleCTL{X}}
\newcommand{\CTLU}{\styleCTL{U}}

\newcommand{\cellHeader}[1]{\cellcolor{blue!20}\textbf{#1}}
\newcommand{\rowHeader}{\rowcolor{blue!20}}

\newcommand{\colCellDec}{\cellcolor{green!40}}
\newcommand{\colCellDecNous}{\cellcolor{green!80}\bfseries}
\newcommand{\colCellUndec}{\cellcolor{red!40}\em}

\newcommand{\colCellUndecNous}{\cellcolor{red!80}\bfseries\em}

\newcommand{\couleurDec}{green}
\newcommand{\couleurUndec}{red}
\newcommand{\couleurOpen}{yellow}

\newcommand{\cellDec}{\cellcolor{\couleurDec}}
\newcommand{\cellUndec}{\cellcolor{\couleurUndec}}
\newcommand{\cellOpenB}{\cellcolor{\couleurOpen}}
\newcommand{\cellOpen}{\cellcolor{\couleurOpen}open}

\definecolor{vertfonce}{rgb}{0.0, 0.5, 0.0}
\definecolor{rougefonce}{rgb}{1, 0.0, 0.0}

\newcommand{\lbbrack}{[{\!}[}%
\newcommand{\rbbrack}{]{\!}]}%
\newcommand{\impl}[1]{\lbbrack #1 \rbbrack}%
\newcommand{\Int}{\texttt{Int}}
\newcommand{\Dist}{\texttt{Dist}}
\newcommand{\rel}{\mathcal{R}}

\newcommand{\st}{\ | \ }
\newcommand{\low}{\texttt{Low}}
\newcommand{\up}{\texttt{Up}}
\newcommand{\suc}{\texttt{Succ}}
\newcommand{\cons}{\texttt{Cons}}

\newcommand{\D}{D}

\newcommand{\hytech}{\textsc{HyTech}}
\newcommand{\imitator}{\textsf{IMITATOR}}
\newcommand{\PSyHCoS}{PSyHCoS}
\newcommand{\romeo}{\textsc{Roméo}}

\newcommand{\symrob}{\texttt{Symrob}}
\newcommand{\uppaal}{\textsc{Uppaal}}

\newcommand{\defProblem}[3]
{%
\noindent\fcolorbox{black}{blue!15}{
	\begin{minipage}{.95\columnwidth}
		\textbf{#1 problem:}\\
		\textsc{Input}: #2\\
		\textsc{Problem}: #3
	\end{minipage}
}
	
	\smallskip
	
}

\newcommand{\mrule}{\;|\;}
\newcommand{\params}{\mathcal{X}}
\newcommand{\PARCTL}{\textnormal{pmARCTL}}
\newcommand{\ARCTL}{\textnormal{ARCTL}}
\newcommand{\mts}{\textnormal{MTS}}
\newcommand{\model}{\mathcal{M}}
\newcommand{\modelSTS}{\mathcal{S}}
\newcommand{\modelSt}{s}
\newcommand{\modelStInit}{{\modelSt^0}}
\newcommand{\modelACTS}{{\mathcal{A}}}
\newcommand{\modelTr}{{\mathcal{T}}}
\newcommand{\modelVS}{{\mathcal{L}}} %
\newcommand{\PV}{\AP} %
\newcommand{\concractset}{B}
\newcommand{\modelPaths}{{\Pi}}
\newcommand{\safe}{safe}
\newcommand{\leftp}{left}
\newcommand{\rightp}{right}
\newcommand{\forwardp}{forw.}
\newcommand{\backp}{back}
\newcommand{\ParVals}{\mathit{ParVals}}
\newcommand{\ActVals}{\ParVals}
\newcommand{\ActVariables}{\params}
\newcommand{\ActSets}{{\mathit{ActSets}}}
\newcommand{\Gom}{{G}}

\newcommand{\concsets}{\ActSets \cup \ActVariables}

\newcommand{\preime}{\textnormal{parPre}^\exists}

\ifdefined\AuthorVersion
	\usepackage[backend=biber,backref=true,style=alphabetic,url=false,doi=true,defernumbers=true,sorting=anyt,maxnames=99]{biblatex} %
	\renewbibmacro*{doi+eprint+url}{%
		\iftoggle{bbx:doi}
			{\color{black!40}\footnotesize\printfield{doi}}
			{}%
		\newunit\newblock
		\iftoggle{bbx:eprint}
			{\usebibmacro{eprint}}
			{}%
		\newunit\newblock
		\iftoggle{bbx:url}
			{\usebibmacro{url+urldate}}
			{}%
	}

\fi
\ifdefined\VeryFinalVersion
\else
	\usepackage[pagewise]{lineno} %
	\linenumbers
	
\fi

\title{Parametric Verification: An Introduction\thanks{%
	\ifdefined\AuthorVersion{}This is the author version of the manuscript of the same name published in the Transactions on Petri Nets and Other Models of Concurrency (\href{https://www.springer.com/gp/computer-science/lncs/lncs-transactions/petri-nets-and-other-models-of-concurrency-topnoc-/731240}{ToPNoC}).
	The published version is available at %
		\href{https://www.springer.com}{\nolinkurl{springer.com}}.
	\fi%
	This work is partially supported by the ANR national research program PACS (ANR-14-CE28-0002).
}}
\author{Étienne André\inst{1}\orcidID{0000-0001-8473-9555}\and Michał Knapik\orcidID{0000-0003-3259-9786} \inst{2}
  \and Didier Lime\orcidID{0000-0001-9429-7586} \inst{3}\and
  Wojciech Penczek\inst{2,4}\orcidID{0000-0001-6477-4863}\and Laure Petrucci\inst{1}\orcidID{0000-0003-3154-5268}}
\institute{
	LIPN, CNRS UMR 7030, Université Paris 13, Villetaneuse, France \and
	Institute of Computer Science, PAS, Warsaw, Poland \and
	École Centrale de Nantes, LS2N, CNRS UMR 6004, Nantes, France \and
	University of Natural Sciences and Humanities, II, Siedlce, Poland
}

\begin{document}
\maketitle

\thispagestyle{plain}

\begin{abstract}
This paper constitutes a short introduction to parametric verification of concurrent
systems. It originates from two 1-day tutorial sessions held at the Petri nets conferences
in Toruń (2016) and Zaragoza (2017). A video of the presentation is available at
\href{https://www.youtube.com/playlist?list=PL9SOLKoGjbeqNcdQVqFpUz7HYqD1fbFIg}{\nolinkurl{youtube.com/playlist?list=PL9SOLKoGjbeqNcdQVqFpUz7HYqD1fbFIg}},
consisting of 14 short sequences.
The paper presents not only the basic formal concepts tackled in the video version, but 
also an extensive literature to provide the reader with further references covering the area.

We first introduce motivation behind parametric verification in general, and then
focus on different models and approaches, for verifying several kinds
of systems.
They include Parametric Timed Automata, for modelling real-time systems, where the
timing constraints are not necessarily known \emph{a priori}. Similarly, Parametric
Interval Markov Chains allow for modelling systems
where probabilities of events occurrences are intervals with parametric bounds.
Parametric Petri Nets allow for compact representation
of systems, and cope with different types of parameters. Finally, Action Synthesis
aims at enabling or disabling actions in a concurrent system to guarantee some of
its properties. 
Some tools implementing these approaches were used during hands-on sessions at the tutorial. 
The corresponding practicals are freely available on the Web.
\end{abstract}

\section{Introduction to parametric verification}
\label{sec:intro}

We first introduce the motivation for performing parametric verification.
Several formal models can be considered, depending on the characteristics of the
system and its properties the designer wants to address. 
We will also discuss the problems of interest in such a framework. At the end of this
section, we will give pointers to some additional and complementary material. 

The reader is assumed to have basic knowledge of Petri nets and/or automata and their
associated verification techniques, since they constitute the basis of the formal
models we address in a parametric setting.

\subsection{Why parameters and of what kind?}
\label{sec:intro:why}
Parameters provide several facilities for easily modelling complex systems:
\begin{enumerate}
\item \emph{Dimensioning}: systems often exhibit components that occur as \emph{multiple
copies} of the same structure or similar ones. For example, a wireless sensor network
is composed of a certain number of identical sensors. At the design phase, the exact number
of sensors might be unknown and would therefore be a parameter.
\item \emph{Choice of actions}: different actions in the system might be possible
in a given state. The designer having several possibilities in mind would model them
and analyse the behaviour of the system with these actions enabled or disabled. 
In this case, the enabledness of each individual action is considered as a parameter.
\item \emph{Design choices}: \emph{different system characteristics} can be taken
into account at the design phase so as to be evaluated. For example, when designing an
electronic system, one might have the choice between different components, available
of course at different prices, each providing its specific characteristics, such as
a better or worse response time. In such a case, the designer may want to construct
a single model, and use these timing characteristics as a parameter so as to evaluate
which is the best possible choice according to his/her needs.
\end{enumerate}

To handle the cases described, different kinds of parameters are used. 
They will influence the type of a formal model and verification techniques chosen. 
In the three previous cases, parameters would be:
\begin{enumerate}
\item \emph{Instances numbering} allow for counting identical components in the system.
\item \emph{Controllable actions} can be enabled or disabled, as opposed
to the other ones which are always available for the system.
\item \emph{Time or probabilities} provide means to handle different characteristics
of the actual system components.
\end{enumerate}

Therefore, many kinds of parameters can be considered according to the problem at hand.

\subsection{Modelling languages}
\label{sec:intro:models}

Among the popular traditional languages for modelling concurrent systems are automata
and Petri nets, and their numerous extensions.

Unfortunately, these modelling languages are not completely suited to handle the systems
of our interest. 
Indeed, numbering instances of components is easily achieved with high-level Petri nets such 
as Coloured Petri Nets (CPNs). 
In CPNs, the Petri net is enriched with data carried by tokens and modified when firing transitions.
These data can very well be a numbering of an instance. Nevertheless, the number of
instances is fixed \emph{a priori}, as opposed to a parameter.
Therefore, when the designer wants to analyse several configurations of the system,
first the model is built for a given number of instances, then analysed, the number
is changed, the model analysed again, and so on.

In Petri nets or automata, there is no specific handling of controllable actions.
Hence, to test several options in the design, each of them must be modelled and analysed
individually.

Finally, timed versions of Petri nets and automata are widely used, but suffer from
the same defaults: values must be known in advance.

Hence, with traditional modelling languages, values are to be set before the analysis
is performed, and the process must be repeated for all possible values. 
It is thus a tedious process which boils down to testing all values one by one, 
taking a huge amount of time.

\subsection{Problems of interest in a parametric setting}
\label{sec:intr:problems}

The major objective for introducing parameters is to circumvent this repetition of analysis,
by introducing parameters in the models. 
Furthermore, the analysis techniques are suited to find constraints on parameters such that desired 
properties are satisfied
(\eg the property is satisfied for all values of the parameter $p$ between 1 and 10) or
even synthesize the set of all such parameters valuations.

The first advantage of this approach is that answering these questions provides all
possible values in a single analysis step. Second, the set of parameters obtained
can be infinite (\eg $p>5$), a result that cannot in general be obtained with an enumerative approach.

\subsection{Sources and references}

Several sources of information are available to the reader, that provide
additional details on the theoretical background, further examples, etc. Among these:
\begin{itemize}
\item the video of the tutorial presentation:
\href{https://www.youtube.com/playlist?list=PL9SOLKoGjbeqNcdQVqFpUz7HYqD1fbFIg}{\nolinkurl{youtube.com/playlist?list=PL9SOLKoGjbeqNcdQVqFpUz7HYqD1fbFIg}};
\item the slides of the tutorial: \href{https://www.imitator.fr/tutorials/PN17/}{\nolinkurl{imitator.fr/tutorials/PN17/}};
\item the exercises with \imitator{} (\href{https://www.imitator.fr/tutorials/PN17/}{\nolinkurl{imitator.fr/tutorials/PN17/}})
and \romeo{} (\href{http://romeo.rts-software.org/doc/tutorial.html}{\nolinkurl{romeo.rts-software.org/doc/tutorial.html}});
\item the extensive literature that is referenced.
\end{itemize}

\paragraph*{Outline.} In \cref{sec:PTA}, we first consider parameters as unknown constants
in timed automata, \ie{} Parametric Timed Automata~\cite{AHV93}.
We review decidability results, and report on decidable subclasses.
Then, in \cref{sec:PIMC}, we consider parameters as unknown probabilities in Parametric
Interval Markov Chains~\cite{BDDLLP-VMCAI16}.
\cref{sec:ppn} deals with parameters that are unknown numbers of tokens in
Petri nets, yielding Parametric Time Petri Nets~\cite{DJLR15,DJLR17}.
As a last formalism, we also consider in \cref{sec:action} the synthesis of actions (seen as Boolean parameters) 
\cite{KnapikMP15,KnapikP14}.
Finally we review some verification tools in \cref{sec:tools}.

\section{Parametric Timed Automata}
\label{sec:PTA}

Classical qualitative model checking, implemented in powerful tools used successfully in industry, 
falls short when \emph{quantitative} aspects of systems such as time, energy, probabilities, etc., are to be verified.
Timed automata~\cite{AD94} allow for modeling and verifying time critical concurrent systems.
This seminal work~\cite{AD94} received the CAV conference award in 2008, and since then 
numerous works have extended the formalism of timed automata.

However, despite of some success, (timed) model checking can be seen as slightly disappointing.
There are two main reason of that:
\begin{enumerate}
	\item the binary response to properties satisfaction may not be informative enough, and
	\item the insufficient abstraction to cater for tuning and scalability of systems.
\end{enumerate}
Adding parameters offers a higher level of abstraction by allowing unknown constants in a model.
Parameters can be used to model unknown \emph{timing} constants of timed systems.
This approach has the following advantages:
\begin{itemize}
\item it becomes possible to verify a system at an earlier design stage, 
       when not all timing constants are known with full certainty.
\item it allows designers to cope with uncertainty even at runtime: some timings constants (\eg{} periods of a real-time system) 
       may be known up to a given precision only (\eg{} given with an interval of confidence), and parameters can model this imprecision.
\end{itemize}
Parametric timed automata~\cite{AHV93} are an extension of timed automata where timing constants can become unknown, \ie{} \emph{parameters}.
They represent a particularly expressive formalism: in fact, its expressiveness is Turing-complete~\cite{ALR16ICFEM} and all non-trivial problems
related to parametric timed automata are undecidable.
For example, the mere existence of a parameter valuation for which there exists a run reaching some location is undecidable
	(see \eg{} \cite{Andre19STTT} for a survey).

Parametric timed automata suffer from negative decidability results, but they still remain a quite powerful formalism.
They can be used to address robustness (in the sense of possibly infinitesimal variations of timing constants~\cite{BMS13}), 
to model and verify systems with uncertain constants, and 
to synthesize suitable (possibly unknown) valuations so that the system meets its specification.

In addition, several recent decidability results for subclasses of parametric timed automata (\eg{} \cite{BlT09,BO14,JLR15,BBLS15,ALime17}) 
has made this formalism more promising, while new algorithmic and heuristic techniques (\eg{} \cite{KP12,JLR15,ALNS15,ABBCR16,LSGA17,ABPV19}) 
has made the parametric verification for some classes of problems more scalable and complete, or more often terminating.

Verification with parametric timed automata has had important outcomes in various areas, with verification of case studies such as
the root contention protocol~\cite{HRSV02}, 
Philip's bounded retransmission protocol~\cite{HRSV02},
a 4-phase handshake protocol~\cite{KP12},
the alternating bit protocol~\cite{JLR15},
an asynchronous circuit commercialised by ST-Microelectronics~\cite{CEFX09},
(non-preemptive) schedulability problems~\cite{JLR15},
a distributed prospective architecture for the flight control system of the next generation of spacecrafts designed at ASTRIUM Space Transportation~\cite{FLMS12},
and even analysis of music scores~\cite{FJ13}.

In this section, we recall the syntax and semantics of parametric timed automata (\cref{ss:syntax-semantics}) and their subclasses (\cref{ss:subclasses}).
We introduce theoretical problems of interest (\cref{section:problems}), and review decidability results (\cref{section:decidability}).

\subsection{Basic notions}
Let %
$\grandn$, $\grandz$, $\grandqplus$, and $\grandrplus$ denote the sets 
of %
non-negative integers, integers, non-negative rational numbers, 
and non-negative real numbers, respectively.

We first define the notions necessary to deal with clocks.
We begin with clock valuations.
\begin{definition}[clock valuation]
Let~$\Clock = \{ \clock_1, \dots, \clock_\ClockCard \} $ be a finite set of \emph{clocks},
\ie{} real-valued variables that evolve at the same rate.

A \emph{clock valuation} is a function $\clockval : \Clock \rightarrow \grandrplus$.

We identify a clock valuation~$\clockval$ with the \emph{point} $(\clockval(\clock_1), \dots, \clockval(\clock_{\ClockCard}))$.
\end{definition}

The following notations allow for specifying null clocks, and adding simultaneously
the same delay to all clocks.

\begin{notation}[clock operations]\mbox{}
\begin{itemize}
\item We write $\Clock = \ClocksZero$ for $\bigwedge_{1 \leq i \leq \ClockCard}\clock_i = 0$.
\item We also use a special zero-clock $\clock_0$, always equal to~0 (as in \eg{} \cite{HRSV02}).
\item Given $d \in \grandrplus$, $\clockval + d$ denotes the valuation such that 
      $(\clockval + d)(\clock) = \clockval(\clock) + d$, for all $\clock \in \Clock$.
	\item Given $\resets \subseteq \Clock$, we define the \emph{reset} of a valuation~$\clockval$, denoted by $\reset{\clockval}{\resets}$, as follows: $\reset{\clockval}{\resets}(\clock) = 0$ if $\clock \in \resets$, and $\reset{\clockval}{\resets}(\clock)=\clockval(\clock)$ otherwise.
\end{itemize}
\end{notation}

The systems considered comprise \emph{a priori} unknown timing constants
that are thus parameters to be synthesized according to the targeted property.

\begin{definition}[timing parameter valuation]
Let $\Param = \{ \param_1, \dots, \param_\ParamCard \}$ be a set of \emph{timing parameters}, 
\ie{} unknown timing constants.

A \emph{timing parameter valuation} $\pval$ is a function
$\pval : \Param \rightarrow \grandqplus$.

We identify a valuation~$\pval$ with the \emph{point} $(\pval(\param_1), \dots, \pval(\param_{\ParamCard}))$.
\end{definition}

Clocks and parameters are used together and thus can be combined.

\begin{notation}[clocks and parameters valuations combined]
Given a timing parameter valuation $\pval$ and a clock valuation $\clockval$,
we denote by $\wv{\clockval}{\pval}$ the valuation over $\Clock\cup\Param$ such that 
for all clocks $\clock$, $\valuate{\clock}{\wv{\clockval}{\pval}}=\valuate{\clock}{\clockval}$
and 
for all timing parameters $\param$, $\valuate{\param}{\wv{\clockval}{\pval}}=\valuate{\param}{\pval}$.
\end{notation}

The expressions on clocks can concern clock themselves, but also involve parameters
and constant delays.

\begin{notation}[linear terms]\mbox{}
\begin{itemize}
\item In the following, let $\lterm$ denote a linear term over $\Clock \cup \Param$
of the form $\sum_{1 \leq i \leq \ClockCard} \gamma_i \clock_i + \sum_{1 \leq j \leq \ParamCard} \beta_j \param_j + d$, with
	$\clock_i \in \Clock$,
	$\param_j \in \Param$,
	and
	$\gamma_i, \beta_j, d \in \grandz$.
\item Let $\plterm$ denote a parametric linear term over $\Param$, that is a linear
term without clocks (i.e., $\gamma_i = 0$ for all $i$).
\end{itemize}
\end{notation}

The synthesis of parameters leads to expressing \emph{constraints} on their values 
in order to guarantee that the model satisfies the expected properties.

\begin{definition}[constraints on clocks and timing parameters]
A \emph{constraint} $\C$ over $\Clock \cup \Param$ is defined by
the following grammar:
\[\phi :=
           \phi \land \phi
	\mrule \neg \phi
	\mrule \lterm \bowtie 0 \text{,}
\]
where ${\bowtie} \in \{<, \leq, \geq, >\}$, $\lterm$ is a linear term.
\end{definition}

\begin{definition}[constraint satisfaction]
A valuation $\wv{\clockval}{\pval}$ satisfies a constraint $\C$, denoted $\wv{\clockval}{\pval} \models
\C$, if the expression obtained by replacing in~$\C$ each timing
parameter by its valuation as in $\pval$ evaluates to $\BTrue$.
\end{definition}

Zones allow for defining convex sets of clocks and timing parameters values.

\begin{definition}[zones and parametric guards]
A \emph{zone}~$\C$ is a constraint
such that each of its linear conjuncts can be written in the form
$\clock_i - \clock_j \bowtie \plterm$, where $\clock_i,\clock_j \in \Clock \cup \{ \clock_0 \}$.
A \emph{parametric guard}~$\guard$ is a zone such that each of its linear conjuncts can be written in the form
$\clock_i \bowtie \plterm$.
\end{definition}

\begin{definition}[satisfiability]
Given a zone $\C$, $\wv{\clockval}{\pval} \models \C$ indicates that
valuating each clock variable~$\clock$ with $\clockval(\clock)$ and each timing parameter~$\param$
with $\pval(\param)$ within $\C$, evaluates to true.
Zone $\C$ is \emph{satisfiable} if $\exists \clockval, \pval \text{ s.t.\ } \wv
{\clockval}{\pval} \models \C$.
\end{definition}

Time elapsing can be obtained by adding a new variable to all clocks, ensuring that
this variable is non-negative, and eliminating it (see, \eg{} \cite{ACEF09}).

\begin{definition}[time elapsing of a zone]
The \emph{time elapsing} of a zone~$\C$, denoted by $\timelapse{\C}$, is the constraint
over $\Clock\cup\Param$ obtained from~$\C$ by delaying all clocks by any arbitrary
amount of time.

That is,
\[\wv{\clockval'}{\pval} \models \timelapse{\C} \text{ iff } \exists \clockval : \Clock \to \grandrplus, \exists d \in \grandrplus \text { s.t. } \wv{\clockval'}{\pval} \models \C \land \clockval' = \clockval + d \text{.}\]

\end{definition}

\subsection{Syntax and semantics}\label{ss:syntax-semantics}
\subsubsection{Syntax}
\begin{definition}[parametric timed automaton \cite{AHV93}]\label{def:PTA}
	A \emph{parametric timed automaton} (PTA) is a tuple \mbox{$\A = (\Actions, \Loc, \locinit, \LocFinal, \Clock, \Param, \invariant, \Edges)$}, where:
	\begin{enumerate}%
		\item $\Actions$ is a finite set of actions,
		\item $\Loc$ is a finite set of locations,
		\item $\locinit \in \Loc$ is the initial location,
		\item $\LocFinal \subseteq \Loc$ is a set of final or accepting locations,
		\item $\Clock$ is a finite set of clocks,
		\item $\Param$ is a finite set of parameters,
		\item $\invariant$ is the invariant, assigning to every $\loc\in \Loc$ a parametric guard $\invariant(\loc)$,
		\item $\Edges$ is a finite set of edges  $\edge = (\loc,\guard,\action,\resets,\loc')$
		where
		$\loc,\loc'\in \Loc$ are the source and target locations, $\action \in \Actions$, $\resets\subseteq \Clock$ is a
		set of clocks to be reset, and
		$\guard$ is a parametric guard called the transition guard.
	\end{enumerate}
\end{definition}

Given a parameter valuation $\pval$, we denote by $\valuate{\A}{\pval}$ the non-parametric timed automaton where all occurrences of each parameter~$\param_i$ have been replaced by~$\pval(\param_i)$.
If $\valuate{\A}{\pval}$ is such that all constants in guards and resets are integers, then $\valuate{\A}{\pval}$ is a \emph{timed automaton}~\cite{AD94}.
In the following, we refer to any structure $\valuate{\A}{\pval}$ as a timed automaton, by assuming a rescaling
of the constants: by multiplying all constants in $\valuate{\A}{\pval}$ by the least common multiple of their denominators, we obtain an equivalent (integer-valued) timed automaton.

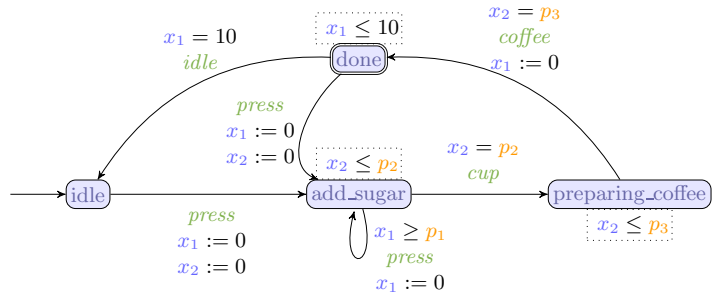
\begin{figure*}%
	\newcommand{\ratio}{0.5\textwidth}
 
	\centering

	\scalebox{.9}{
	\begin{tikzpicture}[scale=2, auto, ->, >=stealth']
 
		\node[location, initial] at (0,0) (idle) {\styleloc{idle}};
 
		\node[location] at (2,0) (add_sugar) {\styleloc{add\_sugar}};
		\node [invariant,above] at (add_sugar.north) {\begin{tabular}{@{} c @{\ } c@{} }& $ \styleclock{x_2} \leq \styleparam{\param_2}$\\\end{tabular}};
 
		\node[location] at (4,0) (preparing_coffee) {\styleloc{preparing\_coffee}};
		\node [invariant,below] at (preparing_coffee.south) {\begin{tabular}{@{} c @{\ } c@{} }& $ \styleclock{x_2} \leq \styleparam{\param_3}$\\\end{tabular}};
 
		\node[location, final] at (2,1) (done) {\styleloc{done}};
		\node [invariant,above] at (done.north) {\begin{tabular}{@{} c @{\ } c@{} }& $ \styleclock{x_1} \leq 10$\\\end{tabular}};

		\path (idle) edge node[below]{\begin{tabular}{@{} c @{\ } c@{} }
		 & $\styleact{press}$\\
		 & $\styleclock{x_1}:=0$\\
		 & $\styleclock{x_2}:=0$\\
		\end{tabular}} (add_sugar);

		\path (add_sugar) edge[loop below] node[right]{\begin{tabular}{@{} c @{\ } c@{} }
		& $ \styleclock{x_1} \geq \styleparam{\param_1}$\\
		 & $\styleact{press}$\\
		 & $\styleclock{x_1}:=0$\\
		\end{tabular}} (add_sugar);

		\path (add_sugar) edge node[above]{\begin{tabular}{@{} c @{\ } c@{} }
		& $ \styleclock{x_2} = \styleparam{\param_2}$\\
		 & $\styleact{cup}$\\
		\end{tabular}} (preparing_coffee);

		\path (preparing_coffee) edge[bend right] node[above]{\begin{tabular}{@{} c @{\ } c@{} }
		& $ \styleclock{x_2} = \styleparam{\param_3}$\\
		 & $\styleact{coffee}$\\
		 & $\styleclock{x_1}:=0$\\
		\end{tabular}} (done);

		\path (done) edge[out=220,in=160] node[left]{\begin{tabular}{@{} c @{\ } c@{} }
		 & $\styleact{press}$\\
		 & $\styleclock{x_1}:=0$\\
		 & $\styleclock{x_2}:=0$\\
		\end{tabular}} (add_sugar);
 
		\path (done) edge[bend right] node[above]{\begin{tabular}{@{} c @{\ } c@{} }
		& $ \styleclock{x_1} = 10$\\
		 & $\styleact{idle}$\\
		\end{tabular}} (idle);
	\end{tikzpicture}
	}
	\caption{A coffee machine modeled using a PTA}
	\label{fig:coffee}
\end{figure*}

\begin{example}
	Consider the coffee machine in \cref{fig:coffee}, modelled using a PTA with 4 locations, 2 clocks ($\styleclock{x_1}$ and $\styleclock{x_2}$) and 3 parameters ($\styleparam{\param_1}, \styleparam{\param_2}, \styleparam{\param_3}$).
	Invariants are boxed.
	The only accepting location (with a double border) is \styleloc{done}.
	Observe that all guards and invariants are simple constraints.
	
	The machine can initially be idle for an arbitrarily long time.
	Then, whenever the user presses the (unique) button (action \styleact{press}),
	the PTA enters location \styleloc{add\_sugar}, resetting both clocks.
	The machine can remain in this location as long as the invariant ($\styleclock{x_2} \leq \styleparam{\param_2}$) is satisfied;
	there, the user can add a dose of sugar by pressing the button (action \styleact{press}), provided the guard ($\styleclock{x_1} \geq \styleparam{\param_1}$) is satisfied, which resets~$\styleclock{x_1}$.
	That is, the user cannot press twice the button (and hence add two doses of sugar) within a time less than~$\styleparam{\param_1}$.
	Then, $\styleparam{\param_2}$ time units after the machine left the idle mode, a cup is delivered (action \styleact{cup}), and the coffee is being prepared;
	eventually, $\styleparam{\param_2}$ time units after the machine left the idle mode, the coffee (action \styleact{coffee}) is delivered.
	Then, after 10 time units, the machine returns to the idle mode---unless a user again requests a coffee by pressing the button.
\end{example}

\subsubsection{Concrete Semantics}
\begin{definition}[Concrete semantics of a TA]
	Given a PTA $\A = (\Actions, \Loc, \locinit, \LocFinal, \Clock, \Param, \invariant, \Edges)$,
	and a parameter valuation~\(\pval\),
	the concrete semantics of $\valuate{\A}{\pval}$ is given by the timed transition system $(\States, \sinit, \flecheRel)$, with
	\begin{itemize}
		\item $\States = \{ (\loc, \clockval) \in \Loc \times \grandrplus^\ClockCard \mid \wv{\clockval}{\pval} \models \invariant(\loc) \}$, 
		\item $\sinit = (\locinit, \ClocksZero) $, and
		\item $\flecheRel$ consists of the discrete and (continuous) delay transition relations:
		\begin{itemize}
			\item discrete transitions: $(\loc,\clockval) \fleche{\edge} (\loc',\clockval')$, %
				if $(\loc, \clockval) , (\loc',\clockval') \in \States$, there exists $\edge = (\loc,\guard,\action,\resets,\loc') \in \Edges$, $\clockval'= \reset{\clockval}{\resets}$, and $\wv{\clockval}{\pval} \models \guard$.
			\item delay transitions: $(\loc,\clockval) \fleche{d} (\loc, \clockval+d)$, with $d \in \grandrplus$, if $\forall d' \in [0, d], (\loc, \clockval+d') \in \States$.
		\end{itemize}
	\end{itemize}
\end{definition}

Moreover we write $(\loc, \clockval)\longuefleche{\edge} (\loc',\clockval')$ for a combination of a delay and discrete transition where
	$((\loc, \clockval), \edge, (\loc', \clockval')) \in \longueflecheRel$ if
		$\exists d, \clockval'' :  (\loc,\clockval) \fleche{d} (\loc,\clockval'') \fleche{\edge} (\loc',\clockval')$.

Given a TA~$\valuate{\A}{\pval}$ with concrete semantics $(\States, \sinit, \flecheRel)$,
we refer to the states of~$\States$ as the \emph{concrete states} of~$\valuate{\A}{\pval}$.
A (concrete) \emph{run} of~$\valuate{\A}{\pval}$ is a possibly infinite alternating sequence of concrete states of $\valuate{\A}{\pval}$ and edges starting from the initial concrete state $\sinit$ of the form 
$\sinit \longuefleche{\edge_0} \state_1\longuefleche {\edge_1} \cdots \longuefleche{\edge_{m-1}} \state_m \longuefleche{\edge_{m}} \cdots$, such that for all 
$i = 0, 1, \dots$: $\edge_i \in \Edges$, and $(\state_i , \edge_i , \state_{i+1})
\in \longueflecheRel$.
Given a state~$\state=(\loc, \clockval)$, we say that $\state$ is reachable (or that $\valuate{\A}{\pval}$ reaches $\state$) if $\state$ belongs to a run of $\valuate{\A}{\pval}$.
By extension, we say that $\loc$ is reachable in~$\valuate{\A}{\pval}$, if there exists a state $(\loc,\clockval)$ that is reachable.
By extension, given a set of locations~$\somelocs \subseteq \Loc$ ($\somelocs$ stands for ``target''), we say that $\somelocs$ is reachable in~$\valuate{\A}{\pval}$, if there exists a location $\loc \in \somelocs$ that is reachable in~$\valuate{\A}{\pval}$.
Given a set of locations~$\somelocs \subseteq \Loc$, we say that a run \emph{stays} in~$\somelocs$ if all of its states $(\loc, \clockval)$ are such that $\loc \in \somelocs$.

A \emph{maximal run} is a run that is either infinite (\ie{} contains an infinite number of discrete transitions), or that cannot be extended by a discrete transition.
A maximal run is \emph{deadlocked} if it is finite, \ie{} contains a finite number of discrete transitions.
By extension, we say that a TA is deadlocked if it contains at least one deadlocked run.

\begin{example}
	Consider again the PTA modeling a coffee machine in \cref{fig:coffee}.
	Let~$\pval$ be the parameter valuation such that $\pval(\parami{1}) = 1$, $\pval(\parami{2}) = 5$ and $\pval(\parami{3}) = 8$.
	
	Given a clock valuation $\clockval$, we denote it by $(\clockval(\clocki{1}), \clockval(\clocki{2}))$.
	For example, $(0, 4.2)$ denotes that $\clockval(\clocki{1}) = 0$ and $\clockval(\clocki{2}) = 4.2$.
	
	The following sequence is a concrete run of $\valuate{\A}{\pval}$.

	\noindent$
	\big(\styleloc{idle}, (0, 0)\big)
		\longuefleche{\styleact{press}}
	\big(\styleloc{add\_sugar}, (0, 0)\big)
		\longuefleche{\styleact{press}}
	\big(\styleloc{add\_sugar}, (0, 1.78)\big)
		\longuefleche{\styleact{press}}
	\big(\styleloc{add\_sugar}, (0, 4.2)\big)
		\longuefleche{\styleact{cup}}
	\big(\styleloc{preparing\_coffee}, (0.8, 5)\big)
		\longuefleche{\styleact{coffee}}
	\big(\styleloc{done}, (0, 8)\big)
		\longuefleche{\styleact{press}}
	\big(\styleloc{add\_sugar}, (0, 0)\big)
	$
	
	As an abuse of notation, we write above each arrow the action name (instead of the edge), as edges are unnamed in \cref{fig:coffee}.
	
	This concrete run is not maximal (it could be extended).
\end{example}
\subsubsection{Language of timed automata}

Let $(\loc_0, \clockval_0) \longuefleche{\edge_0} (\loc_1, \clockval_1) \longuefleche {\edge_1} \cdots \longuefleche{\edge_{m-1}} (\loc_m, \clockval_m) \longuefleche{\edge_{m}} \cdots$ be a (finite or infinite) run of a TA~$\valuate{\A}{\pval}$.
The associated \emph{untimed word} is $\action_0 \action_1 \cdots \action_m \cdots$, where $\action_i$ is the action of edge~$\edge_i$, for all $i \geq 0$;
the associated \emph{trace}\footnote{%
	This is a non-standard definition of traces (compared to \eg{} \cite{vanGlabbeek90}), but we keep this term as it is used in \eg{} \cite{ACEF09,AM15}.
} is $\loc_0 \action_0 \loc_1 \action_1 \loc_2 \cdots \action_m \loc_{m+1} \cdots$

Given a run $(\loc_0, \clockval_0) \longuefleche{\edge_0} (\loc_1, \clockval_1) \longuefleche {\edge_1} \cdots \longuefleche{\edge_{m-1}} (\loc_m, \clockval_m)$, we say that this run is \emph{accepting} if $\loc_m \in \LocFinal$.

We define the untimed language as the set of all untimed words associated with accepting runs of a TA.

\begin{definition}[untimed language of a TA]\label{definition:language}
	Given a PTA $\A = (\Actions, \Loc, \locinit, \LocFinal, \Clock, \Param, \invariant, \Edges)$,
	and a parameter valuation~\(\pval\),
	the \emph{untimed language} of $\valuate{\A}{\pval}$, denoted by~$\UL(\valuate{\A}{\pval})$, is the set of untimed words associated with all accepting runs of~$\valuate{\A}{\pval}$.
\end{definition}

We define the trace set as the set of traces associated with the accepting runs.

\begin{definition}[trace set of a TA]
	Given a PTA $\A = (\Actions, \Loc, \locinit, \LocFinal, \Clock, \Param, \invariant, \Edges)$,
	and a parameter valuation~\(\pval\),
	the \emph{trace set} of $\valuate{\A}{\pval}$ %
		is the set of traces associated with all accepting runs of~$\valuate{\A}{\pval}$.
\end{definition}
\begin{example}
	Consider again the PTA~$\A$ modeling a coffee machine in \cref{fig:coffee}.
	Let~$\pval$ be the parameter valuation such that $\pval(\parami{1}) = 1$, $\pval(\parami{2}) = 5$ and $\pval(\parami{3}) = 8$.
	
	The untimed language of~$\valuate{\A}{\pval}$ can be described as follows:
	\[\styleact{press}^{[1..6]}\ \styleact{cup}\ \styleact{coffee}
			\big(
				\styleact{idle}^?\ \styleact{press}^{[1..6]}\ \styleact{cup}\ \styleact{coffee}
			\big)^*
	\]
	where $\styleact{\action}^{[a,b]}$, $\styleact{\action}^?$, $\styleact{\action}^*$ denote between $a$ and~$b$ occurrences, zero or one occurrence, and zero or more occurrence(s) of~$\styleact{\action}$, respectively. %

	The trace set of~$\valuate{\A}{\pval}$ can be described as follows:
	\begin{align*}
		\styleloc{idle}\ (\styleact{press}\ \styleloc{add\_sugar})^{[1..6]}\ \styleact{cup}\
		\styleloc{preparing\_coffee}\ \styleact{coffee}\ \styleloc{done}\\
		\big(
				(\styleact{idle} \ \styleloc{idle})^? \ (\styleact{press}\ \styleloc{add\_sugar})^
				{[1..6]}\ \styleact{cup}\ \styleloc{preparing\_coffee}\ \styleact{coffee}\ \styleloc{done}
			\big)^*
	\end{align*}

\end{example}
\subsubsection{Symbolic semantics}

Let us now recall the symbolic semantics of PTAs (see \eg{} \cite{HRSV02,ACEF09,JLR15}).

\begin{definition}[Symbolic state]
	A symbolic state is a pair $(\loc, \C)$ where $\loc \in \Loc$ is a location, and $\C$ its associated parametric zone.
\end{definition}
\begin{definition}[Symbolic semantics]\label{def:PTA:symbolic}
	Given a PTA $\A = (\Actions, \Loc, \locinit, \LocFinal, \Clock, \Param, \invariant, \Edges)$,
	the symbolic semantics of~$\A$ is the labelled transition system called \emph{parametric zone graph}
	$ \PZG = ( \Edges, \SymbState, \symbstateinit, \symbtrans )$, with
	\begin{itemize}
		\item $\SymbState = \{ (\loc, \C) \mid \C \subseteq \invariant(\loc) \}$, %
		\item $\symbstateinit = \big(\locinit, \timelapse{(\bigwedge_{1 \leq i\leq\ClockCard}\clock_i=0)} \land \invariant(\loc_0) \big)$,
				and
		\item $\big((\loc, \C), \edge, (\loc', \C')\big) \in \symbtrans $ if $\edge = (\loc,\guard,\action,\resets,\loc')$ and
			\[\C' = \timelapse{\big(\reset{(\C \land \guard)}{\resets}\land \invariant(\loc')\big )} \land \invariant(\loc')\]
			with $\C'$ satisfiable.
	\end{itemize}

\end{definition}

That is, in the parametric zone graph, nodes are symbolic states, and arcs are labelled by \emph{edges} of the original PTA.

If $\big((\loc, \C), \edge, (\loc', \C')\big) \in \symbtrans $, we write $\Succ(\symbstate, \edge) = (\loc', \C')$.

\begin{figure}
	\definecolor{MyLoc}{rgb}{0.5,0.1,0.5}
	\definecolor{MyGreen}{rgb}{0.1,0.5,0.1}
	\definecolor{MyValu}{rgb}{0.4,0.3,0.0}

{\centering

	\begin{tikzpicture}[scale=0.3]

	\draw[->,color=black!40] (-2,-2) -- (25, -2);
	\draw[->,color=black!40] (-2,-2) -- (-2, 12);

	\draw[color=MyValu,fill=MyValu!30] (0,0) -- (5,0) -- (10, 5) -- (10, 8) -- (3, 8) -- (0,5) -- (0,0);
	\draw (3, 4) node {$\C$};

	\draw[color=MyGreen,thick] (6,6) -- (10,6) -- (10,8) -- (6,8) -- (6,6);
	\draw(9,9) node (G) {$\guard$};

	\draw[color=MyGreen,fill=MyGreen!30] (6,6) -- (10,6) -- (10,8) -- (6,8) -- (6,6);

	\draw[draw=none, fill=MyValu!10] (7, -2) -- (13, 4) -- (17, 8) -- (18.5, 6.5) -- (14, 2) -- (10, -2);
	\draw[color=MyValu, dashed] (13, 4) -- (18, 9);
	\draw[color=MyValu, dashed] (14, 2) -- (20, 8);

	\draw[thick,color=MyLoc] (10, -2) -- (14, -2) -- (14, 4) -- (7, 4) -- (7, -2);    
	\draw(16, -1) node {$\invariant(\loc')$};

	\draw[color=MyValu, fill=MyValu!30] (7, -2) -- (13, 4) -- (14, 4) -- (14, 2) -- (10, -2);
	\draw(12, 1.5) node (MU) {$\C'$};

	\draw[very thick, color=MyGreen] (6, -2) -- (10, -2);
	\draw[very thick, ->, color=MyGreen] (8, 7) -- (8, -2);
	\draw (5, -2) node {$\resets$};

	\end{tikzpicture}
	
	}
	
	\caption{Computing the successor of a symbolic state}
	\label{figure:succ}
\end{figure}
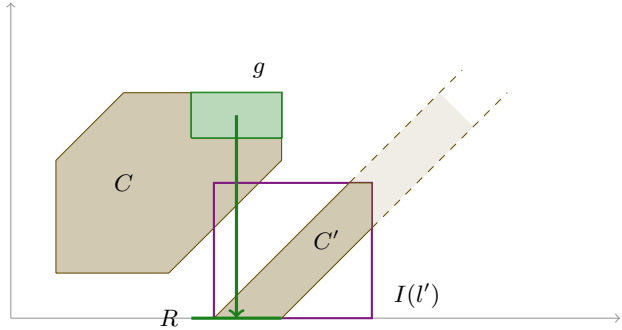

A graphical illustration of the computation of $\Succ$ is given in \cref{figure:succ}.\footnote{%
	This figure comes from~\cite{AS13}, itself coming from an adaptation of a figure by Ulrich~Kühne.
} Starting from the parametric zone $\C$, it is intersected with guard $\guard$, leading
to the parametric values that allow for taking the transition. Then the necessary
clocks are reset. However, for the transition to be taken, the new values thus obtained
must satisfy the invariant of the target location, $\invariant(\loc')$. After this,
when in $\loc'$, time can elapse as long as the invariant still holds, leading to
the new zone $\C'$.

A symbolic run of a PTA is an alternating sequence of symbolic states and edges starting from the initial symbolic state, of the form 
$\symbstate_0 \Fleche{\edge_0} \symbstate_1\Fleche {\edge_1} \cdots \Fleche{\edge_{m-1}} \symbstate_m$, such that for all $i = 0, \dots, m-1$, $\edge_i \in \Edges$, $\symbstate_i, \symbstate_{i+1} \in \SymbState$ and %
$(\symbstate_i, \edge, \symbstate_{i+1}) \in \symbtrans$.
Given a symbolic state~$\symbstate$, we say that $\symbstate$ is reachable if $\symbstate$ belongs to a symbolic run of~$\A$.
In the following, we simply refer to symbolic states belonging to a run of~$\A$ as symbolic states %
of~$\A$.

\begin{example}
	Consider again the coffee machine example in \cref{fig:coffee}.
	A (non-maximal) symbolic run is as follows:

	\noindent$\big(\styleloc{idle}, \clocki{1} = \clocki{2} \land \clocki{1} \geq 0 \big)
		\Fleche{\styleact{press}}
	\big(\styleloc{add\_sugar}, \clocki{1} = \clocki{2} \land 0 \leq \clocki{2} \leq
	\parami{2} \big)$

		$\Fleche{\styleact{press}}
	\big(\styleloc{add\_sugar}, \parami{1} \leq \clocki{2} - \clocki{1} \leq \parami{2}
	\land 0 \leq \clocki{2} \leq \parami{2} \big)$

		$\Fleche{\styleact{press}}
	\big(\styleloc{add\_sugar}, 2 \times \parami{1} \leq \clocki{2} - \clocki{1} \leq \parami{2} \land 0 \leq \clocki{2} \leq \parami{2} \big)$

		$\Fleche{\styleact{cup}}
	\big(\styleloc{preparing\_coffee}, 2 \times \parami{1} \leq \clocki{2} - \clocki{1} \leq \parami{2} \land \parami{2} \leq \clocki{2} \leq \parami{3} \big)$

		$\Fleche{\styleact{coffee}}
	\big(\styleloc{done}, 0 \leq \clocki{1} \leq 10 \land \clocki{2} - \clocki{1}
	= %
		\parami{3}
		\land 2 \times \parami{1} \leq \parami{2} \leq \parami{3}
	\big)$
		
		$\Fleche{\styleact{press}}
	\big(\styleloc{add\_sugar}, \clocki{1} = \clocki{2} \land 0 \leq \clocki{2} \leq
	\parami{2} \land 2 \times \parami{1} \leq \parami{2} \leq \parami{3} \big)
	$
	
	(For sake of readability, we use action names instead of edges along the transitions.)
	
	The parametric zone graph of this example is infinite.
\end{example}
Given a concrete (respectively symbolic) run $(\locinit, \vec{0}) \longuefleche{\edge_0} (\loc_1, \clockval_1) \longuefleche {\edge_1} \cdots \longuefleche{\edge_{m-1}} (\loc_m, \clockval_m)$ (respectively  $(\loc_0, \C_0) \Fleche{\edge_0} (\loc_1, \C_1) \Fleche {\edge_1} \cdots \Fleche{\edge_{m-1}} (\loc_m, \C_m)$),
we define the corresponding discrete sequence as
$\loc_0  \Fleche{\edge_0} \loc_1 \Fleche {\edge_1} \cdots \Fleche{\edge_{m-1}} \loc_m $.
Two runs (concrete or symbolic) are said to be \emph{equivalent} if their associated discrete sequences are equal.

Two important results (see \eg{} \cite{HRSV02}) relate the concrete and the symbolic
semantics, and are recalled below using our syntax.
They provide a sort of equivalence between symbolic parametric zones and concrete runs.
That is, they guarantee that the zones correspond to feasible runs (correctness) and that each run is represented by a zone (completeness).

\begin{lemma}[{\cite[Proposition 3.17]{HRSV02}}]
	\label{Lemma:HRSV02:Prop3.17}
    For each parameter valuation~$\pval$ and clock valuation $\clockval$, if there is a symbolic run in~$\A$ reaching state $(\loc, \C)$, with $\clockval \models \valuate{\C}{\pval}$, then there is an equivalent concrete run in $\valuate{\A}{\pval}$ reaching state $(\loc, \clockval)$.
\end{lemma}
\begin{lemma}[{\cite[Proposition 3.18]{HRSV02}}]
	\label{Lemma:HRSV02:Prop3.18}
    For each parameter valuation~$\pval$ and clock valuation $\clockval$, if there is a concrete run in $\valuate{\A}{\pval}$ reaching state $(\loc, \clockval)$, then there is an equivalent symbolic run in~$\A$ reaching a state $(\loc, \C)$ such that $\clockval \models \valuate{\C}{\pval}$.
\end{lemma}
\subsection{Subclasses of PTAs}\label{ss:subclasses}

Lower-bound/upper-bound parametric timed automata (L/U-PTAs), proposed in~\cite{HRSV02}, restrict the use of parameters in the model.
\begin{definition}[L/U-PTA]\label{def:LUPTA} %
	An L/U-PTA is a PTA where the set of parameters is partitioned into lower-bound parameters and upper-bound parameters,
	where an upper-bound (resp.\ lower-bound) parameter~$\param_i$ is such that, 
    for every guard or invariant constraint $\clock \compOp \sum_{1 \leq j \leq \ParamCard} \beta_j \param_j + d$, we have:
		$\beta_j > 0$ implies ${\compOp} \in \{ \leq, < \}$ (resp.\ ${\compOp} \in \{ \geq, > \}$),
		and
		$\beta_j < 0$ implies ${\compOp} \in \{ \geq, > \}$ (resp.\ ${\compOp} \in \{ \leq, < \}$).
\end{definition}
In~\cite{BlT09}, two additional subclasses are introduced: L-PTAs (resp.\ U-PTAs) are PTAs with only lower-bound (resp.\ upper-bound) parameters.

L/U-PTAs enjoy a well-known monotonicity property~\cite{HRSV02}: increasing upper-bound parameters or decreasing lower-bound parameters can only add behaviours.

\begin{example}
	Consider again the coffee machine in \cref{fig:coffee}, modelled using a PTA~$\A$.
	This PTA is not an L/U-PTA; indeed, in the guard $\styleclock{x_2} = \styleparam{\param_2}$ (resp.\ $\styleclock{x_2} = \styleparam{\param_3}$), $\styleparam{\param_2}$ (resp.~$\styleparam{\param_3}$) is compared with clocks both as a lower-bound and as an upper-bound.
	(Recall that $=$ stands for $\leq$ and $\geq$.)
	
	However, if one replaces $\styleclock{x_2} = \styleparam{\param_2}$ with $\styleclock{x_2} \leq \styleparam{\param_2}$
	and $\styleclock{x_2} = \styleparam{\param_3}$ with $\styleclock{x_2} \leq \styleparam{\param_3}$, then $\A$ becomes an L/U-PTA with lower-bound parameter $\styleparam{\param_1}$ and upper-bound parameters $\{\styleparam{\param_2}, \styleparam{\param_3}\}$.
	Note that equalities are not forbidden in L/U-PTAs (\eg{} $\styleclock{x_1} =
	10$), but only equalities involving parameters are.
\end{example}

Several case studies fit into the class of L/U-PTAs: the root contention protocol, the bounded retransmission protocol and the Fischer mutual exclusion protocol are all modelled with L/U-PTAs in~\cite{HRSV02};
in \cite{HRSV02,KP12}, both the Fischer mutual exclusion protocol and a producer-consumer are verified using L/U-PTAs.
Interestingly, the two case studies of the seminal paper on PTAs~\cite{AHV93} (\viz{} a toy train gate controller model and a model of Fischer mutual exclusion protocol) are also L/U-PTAs, although the concept of L/U-PTAs had not yet been proposed at that time.
In addition, most models of asynchronous circuits with bi-bounded delays (\ie{} where each delay between the change of an input signal and the change of the corresponding output is a parametric interval) can be modelled using L/U-PTAs.

We will also consider \emph{bounded} PTAs, \ie{} PTAs with a bounded parameter domain that assigns to each parameter an infimum and a supremum, both integers.

\begin{definition}[bounded PTA]\label{definition:bounded}
	A \emph{bounded PTA} is $\bounded{\A}{\bounds}$, where $\A$ is a PTA, and $\bounds%
	$ assigns to each parameter~$\param$ an interval
	$[\boundinf, \boundsup]$,
		$(\boundinf, \boundsup]$,
		$[\boundinf, \boundsup)$,
		or
		$(\boundinf, \boundsup)$,
	with $\boundinf, \boundsup \in \grandn$.
	We use $\boundinfin{\param}{\bounds}$ and $\boundsupin{\param}{\bounds}$ to denote the infimum and the supremum of~$\param$, respectively.
	(Note that we rule out $\infty$ as a supremum.)
	
	We say that a bounded PTA is a \emph{closed bounded PTA} if, for each parameter~$\param$, its ranging interval $\bounds(\param)$ is of the form $[\boundinf, \boundsup]$; otherwise it is an \emph{open bounded PTA}.
	
	We define similarly bounded L/U-PTAs.
\end{definition}
\subsection{Decision and computation problems}\label{section:problems}
\subsubsection{TCTL}\label{section:TCTL}

TCTL~\cite{ACD93} is the quantitative extension of CTL where temporal modalities are augmented with constraints on duration.
Formulae are interpreted over timed transition systems.

Given~$\atomicprop \in \AP$ and $c\in\grandn$, the language of TCTL %
is given by the following grammar: 
\[
  \formule::= \quad \top\quad|\quad \atomicprop \quad|\quad \neg{\formule}\quad|\quad\formule\wedge\formule\quad |\quad \CTLE{}\formule\CTLU{}_{\compOp c}\formule \quad| \quad \CTLA{}\formule\CTLU{}_{\compOp c}\formule
\]

$\CTLA{}$ reads ``always'',
$\CTLE{}$ reads ``exists'',
and
$\CTLU{}$ reads ``until''.

Standard abbreviations include Boolean operators as well as~$\styleCTL{EF}_{\compOp c}\formule$ 
for $\CTLE{} \top \CTLU{}_{\compOp c}\formule$, $\styleCTL{AF}_{\compOp c}\formule$ 
for $\CTLA{} \top \CTLU{}_{\compOp c}\formule$ and $\styleCTL{EG}_{\compOp c}\formule$ 
for $\neg\styleCTL{AF}_{\compOp c}\neg\formule$.
$\CTLF{}$ reads ``eventually'' while $\CTLG{}$ reads ``globally''.

\begin{definition}[Semantics of TCTL]
Given a TA $\pval(\A)$, the following clauses define when a state $\state_i$ of its timed transition system $(\States, \sinit, \flecheRel)$ 
satisfies a TCTL formula $\formule$, denoted by $\state_i\models\formule$, 
by induction over the structure of $\formule$ (semantics of Boolean operators is omitted:
\begin{enumerate}
	\item $\state_i\models\CTLE{}\formule\CTLU{}_{\compOp c}\psi$ 
	if there is a maximal run~$\rho$ in~$\pval(\A)$ 
	with~$\sigma=\state_i\longuefleche {\edge_i} \cdots \longuefleche{\edge_{j-1}} \state_j$ ($i<j$) 
	a prefix of~$\rho$
	s.t.\ $\state_j\models\psi$,
	$\Time(\sigma)\compOp c$, and $\forall k, i\leq k<j:\state_k\models\formule$
	\item $\state_i\models\CTLA{}\formule\CTLU{}_{\compOp c}\psi$
	if for each maximal run~$\rho$ in~$\pval(\A)$ 
	there exists~$\sigma=\state_i\longuefleche {\edge_i} \cdots \longuefleche{\edge_{j-1}} \state_j$ ($i<j$)
	a prefix of~$\rho$
	s.t.\ $\state_j\models\psi$,
	$\Time(\sigma)\compOp c$, and $\forall k, i\leq k<j:\state_k\models\formule$.
\end{enumerate}
\end{definition}
where, given a concrete run~$\rho$, $\Time(\rho)$ gives the total sum of the delays~$d$ along~$\rho$.

In $\CTLE{}\formule\CTLU{}_{\compOp c}\psi$
the classical until is extended by requiring that $\formule$ be
satisfied within a duration (from the current state) verifying the constraint ``$\compOp c$''.
Given~$\pval$, a PTA $\A$ and a TCTL formula~$\formule$, we write~$\pval(\A) \models \formule$ when~$\sinit \models\formule$.

\subsubsection{Decision problems}

\paragraph{Emptiness and universality of the valuations set.}
Let \Problem{} be a given a class of decision problems (reachability, unavoidability, etc.).

\defProblem
	{\Problem-emptiness}
	{A PTA~\A{} and an instance $\varproblem$ of \Problem{}}
	{Is the set of parameter valuations $\pval$ such that $\valuate{\A}{\pval}$ satisfies $\varproblem$ empty?}

\defProblem
	{\Problem-universality}
	{A PTA~\A{} and an instance $\varproblem$ of \Problem{}}
	{For all parameter valuations $\pval$, does $\valuate{\A}{\pval}$ satisfy $\varproblem$?}

In this section, we mainly focus on the following decision problems:
\begin{itemize}
	\item reachability (\EF{}\footnote{%
		The names ``\EF{}'', ``\AF{}'', ``\EG{}'' come from the TCTL syntax, and are consistent with the notations introduced in~\cite{JLR15} and subsequently used in further papers (such as \cite{ALR16ICFEM,ALime17}).
	}): given a TA $\valuate{\A}{\pval}$, is there at least one run of $\valuate{\A}{\pval}$ that reaches a given location?
		That is, \EF{}-emptiness asks: ``is the set of parameter valuations~$\pval$ such that the TA $\valuate{\A}{\pval}$ reaches a given location empty?''
		And \EF{}-universality asks: ``are all parameter valuations such that the corresponding TA reaches a given location?''
	
	\item unavoidability (\AF{}): given a TA $\valuate{\A}{\pval}$, do all runs of $\valuate{\A}{\pval}$ eventually reach a given location?

	\item \EG{}: given a TA~$\valuate{\A}{\pval}$ and a subset~$\somelocs$ of its locations, is there at least one maximal run of~$\valuate{\A}{\pval}$ that always stays in~$\somelocs$?

	\item \AG{}: given a TA~$\valuate{\A}{\pval}$ and a subset~$\somelocs$ of its locations, do all runs of~$\valuate{\A}{\pval}$ stay in~$\somelocs$?

	\item deadlock-existence (\ED): given a TA $\valuate{\A}{\pval}$, is there at least one maximal run of $\valuate{\A}{\pval}$ that is deadlocked, \ie{} has no discrete successor (possibly after some delay)?
	
	\item cycle-existence (\EC): given a TA~$\valuate{\A}{\pval}$, is there at least one run of~$\valuate{\A}{\pval}$ with an infinite number of discrete transitions?	
\end{itemize}

Note that \AF{}-emptiness is equivalent to \EG{}-universality, while \AG{}-emptiness is equivalent to \EF{}-universality.

We will finally consider the following two additional emptiness problems:

\defProblem
	{Language-preservation-emptiness}
	{A PTA~\A{} and a parameter valuation~$\pval'$}
	{Is the set of parameter valuations $\pval$ such that $\pval \neq \pval'$ and for which $\valuate{\A}{\pval}$ has the same untimed language as $\valuate{\A}{\pval'}$ empty?}

\defProblem
	{Trace-preservation-emptiness}
	{A PTA~\A{} and a parameter valuation~$\pval'$}
	{Is the set of parameter valuations $\pval$ such that $\pval \neq \pval'$ and for which $\valuate{\A}{\pval}$ has the same set of traces as $\valuate{\A}{\pval'}$ empty?}

\subsubsection{Computation problem}

Additionally, we define the following computation problem:

\defProblem
	{\Problem-synthesis}
	{A PTA~\A{} and an instance $\varproblem$ of \Problem{}}
	{Compute the parameter valuations such that $\valuate{\A}{\pval}$ satisfies $\varproblem$.}

\begin{example}
	Let us exemplify some decision and computation problems for the coffee machine
	PTA in \cref{fig:coffee}.
	Assume the unique target location is \styleloc{done}, \ie{} $\somelocs=\{ \styleloc{done}
	\}$.

	\EF{}-emptiness asks whether the set of parameter valuations that can reach location
	\styleloc{done} for some run is empty, \ie{} there is an execution in which the
	coffee is not delivered. This is false (\eg{} $\styleparam{\param_1}
	= 1$, $\styleparam{\param_2} = 2$, $\styleparam{\param_3} = 3$ can reach \styleloc{done}).

	\EF{}-universality asks whether all parameter valuations can reach location \styleloc{done}
	for some run, \ie{} all executions allow for delivering the coffee, regardless
	of the parameters valuation. This is false (no parameter valuation such that $
	\styleparam{\param_2} > \styleparam{\param_3}$ can reach \styleloc{done}).

	\AF{}-emptiness asks whether the set of parameter valuations that can reach location
	\styleloc{done} for all runs is empty, \ie{} if there is some parameters valuation
	for which coffee delivery is not guaranteed. This is false (\eg{} $\styleparam{\param_1}
	= 1$, $\styleparam{\param_2} = 2$, $\styleparam{\param_3} = 3$ cannot avoid \styleloc{done}).

	\EF{}-synthesis consists in synthesizing all valuations for which a run reaches
	location \styleloc{done}, \ie{} identifies all parameters valuations for which
	a coffee will eventually be delivered. The resulting set of valuations is $0 \leq
	\styleparam{\param_2} \leq \styleparam{\param_3} \leq 10 \land \styleparam{\param_1} \geq 0$.
\end{example}

\subsection{Decidability}\label{section:decidability}
\subsubsection{The general class of PTAs}

With the rule of thumb that all problems are undecidable for PTAs, we review the decidability of the aforementioned problems.
\begin{itemize}
	\item $\EF$-emptiness was shown to be undecidable~\cite{AHV93}, with different
	flavours and settings: for a single bounded parameter~\cite{Miller00}, for a single rational-valued or integer-valued parameter~\cite{BBLS15}, with only one clock compared to parameters~\cite{Miller00}, or with strict constraints only~\cite{Doyen07}.
	\item $\AF$-emptiness was shown undecidable in~\cite{JLR15}.
	\item $\AG$-emptiness was shown undecidable in~\cite{ALR16ICFEM}.
	\item $\EG$-emptiness (as well as \EC{} and \ED{}) were shown undecidable in~\cite{ALime17}.
	\item The language and trace-preservation problems were shown undecidable in~\cite{AM15}.
\end{itemize}
A complete survey is available in~\cite{Andre19STTT}.

\paragraph{}
Following the very negative results for PTAs, subclasses have been proposed.
We review some in the following.

\subsubsection{The class of L/U-PTAs}
\paragraph{A main decidability result.}

The first (and main) positive result for L/U-PTAs is the decidability of the \EF{}-emptiness problem~\cite{HRSV02}.
L/U-PTAs benefit from the following interesting monotonicity property: increasing the value of an upper-bound parameter or decreasing the value of a lower-bound parameter necessarily relaxes the guards and invariants, and hence can only add behaviours.
Therefore, checking the \EF{}-emptiness of an L/U-PTA can be achieved by replacing
all lower-bound parameters with~0, and all upper-bound parameters with a sufficiently large constant; this yields a non-parametric TA, for which emptiness is PSPACE-complete~\cite{AD94}.
This procedure is not only sound but also complete.

\paragraph{Undecidability results.}

The first undecidability results for L/U-PTAs are shown in~\cite{BlT09}:
the \emph{constrained} \EF{}-emptiness problem and constrained \EF{}-universality problem (for infinite runs acceptance properties) are undecidable for L/U-PTAs.
By constrained it is meant that some parameters of the L/U-PTA can be constrained
by an initial linear constraint, \eg{} $\styleparam{\param_1} \leq 2 \times \styleparam{\param_2}
+ \styleparam{\param_3}$.
Indeed, using linear constraints, one can constrain an upper-bound parameter to be equal to a lower-bound parameter, and hence build a 2-counter machine using an L/U-PTA.
However, when no upper-bound parameter is compared to a lower-bound parameter (\ie{}
when no initial linear inequality contains both an upper-bound and a lower-bound parameter),
these two problems become decidable~\cite{BlT09}.
The exact decidability frontier may have not been found yet: the case where a lower-bound
parameter is constrained to be less than or equal to an upper-bound parameter fits in none of the considered cases.

A second negative result is shown in~\cite{JLR15}: the \AF{}-emptiness problem is undecidable for L/U-PTAs.
This restricts again the use of L/U-PTAs, as \AF{} is essential to show that all possible runs of a system eventually reach a (good) state.

Third, in~\cite{AM15}, the language- and trace-preservation problems were shown to
be undecidable for L/U-PTAs.

\paragraph{Model-checking L/U-PTAs.}

In~\cite{BlT09}, a parametric extension of the dense-time linear temporal logic \mitlzeroinf{} (denoted ``\pmitlzeroinf{}'') is proposed; when parameters are used only as lower or upper bound in the formula (to which we refer as L/U-\pmitlzeroinf{}), satisfiability and model checking are PSPACE-complete; this is obtained by translating the formula into an L/U-automaton and checking an infinite acceptance property.

Then, in~\cite{GLN15}, an extension of MITL allowing parametric linear expressions in bounds is proposed (yielding PMITL).
Two sets of (integer-valued) parameter valuations are considered: 
\begin{enumerate}
	\item the set of valuations for which a PMITL formula is satisfiable, \ie{} for which there exists a timed sequence (possibly belonging to a given L/U-PTA) satisfying it, and
	\item the set of valuations for which a PMITL formula is valid, \ie{} for which all timed sequences (possibly belonging to a given L/U-PTA) satisfy it.
\end{enumerate}
Under some assumptions, the emptiness and universality of the valuation set for which a PMITL property is satisfiable or valid (possibly \wrt{} a given L/U-PTA) are decidable, and EXPSPACE-complete.
Essential assumptions for decidability include the fact that parameters should be used with the same polarity (positive or negative coefficient, as lower or upper bound in the intervals) within the entire PMITL formula, and each interval can only use parameters in one of the endpoints.
Additional assumptions include that no interval of the PMITL formula should be punctual (nor empty), and linear parametric expressions are only used in right endpoints of the intervals (single parameters can still be used as left endpoints).
In addition, two fragments of PMITL are showed to be in PSPACE, including one that
allows for expressing parameterized response (``if an event occurs, then another event shall occur within some possibly parametric time interval'').

Finally, we showed that the emptiness-problem using \emph{nested} quantifiers (\ie{} beyond \EF, \EG, \AF, \AG) automatically leads to the undecidability, even for the very restricted class of U-PTAs with a single parameter (that can even be integer-valued)~\cite{ALR18FORMATS}.
In other words, the nested TCTL emptiness problem is undecidable for U-PTAs.
We may wonder if the \emph{timed} aspect of TCTL (and notably the urgency required by the TCTL formula $\EG \AG_{= 0}$) is responsible for the undecidability.
In fact, it is not, and we could modify the proof to show that CTL itself leads to undecidability, \ie{} that $\EG \CTLA \CTLX$-emptiness is undecidable.

\paragraph{Intractability of the synthesis.}

A very  disappointing result concerning L/U-PTAs is shown in~\cite{JLR15}:
despite decidability of the underlying decision problems (\EF{}-emptiness and \EF{}-universality),
the solution to the \EF{}-synthesis problem for L/U-PTAs, if it can be computed, cannot be represented using a formalism for which the emptiness of the intersection with equality constraints is decidable.
	The proof relies on the undecidability of the constrained emptiness problem of~\cite{BlT09}.
A very annoying consequence is that such a solution cannot be represented as a finite union of polyhedra (since the emptiness of the intersection with equality constraints is decidable).

\paragraph{Liveness.}

The \EG{}-emptiness problem stands at the frontier between decidability and undecidability for the class of L/U-PTAs:
while this problem is decidable for L/U-PTAs with a bounded parameter domain with closed bounds, it becomes undecidable if either the assumption of boundedness or of closed bounds is lifted~\cite{ALime17}.

The deadlock-existence emptiness problem is undecidable, even for the restricted class of closed bounded L/U-PTAs~\cite{ALime17}.

In contrast to deadlock-freeness that is consistently undecidable, and to \EG{}-emptiness for which the frontier between decidability and undecidability is thin, the existence of a parameter valuation for which there exists at least one infinite run (\EC-emptiness) is consistently decidable for L/U-PTAs~\cite{ALime17}.

\subsubsection{The power of integer points}

Following works related integer clock and parameter valuations with decidability in~\cite{JLR15}, we introduced in~\cite{ALR16ICFEM} \emph{integer-points parametric timed automata} (IP-PTAs for short), \ie{} a subclass of PTAs in which any symbolic state contains at least one integer point.

\begin{definition}\label{definition:IP-PTA}
	A PTA~$\A$ is an \emph{integer points PTA} (in short \emph{IP-PTA}) if, in any reachable symbolic state $(\loc, \C)$ of~$\A$, $\C$ contains at least one integer point, \ie{} $\exists \pval : \Param \to \grandn, \exists \clockval : \Clock \to \grandn \text{ s.t. } \wv{\clockval}{\pval} \models \C$.
\end{definition}
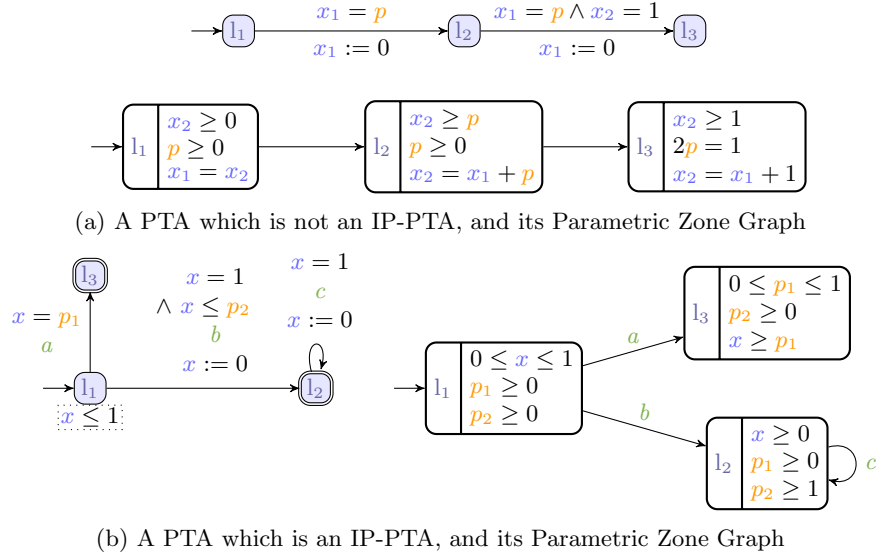
\begin{figure}
	\begin{subfigure}[b]{\textwidth}
		\centering

		\begin{tikzpicture}[scale=1, node distance=3cm, auto, ->, >=stealth']
		
		\footnotesize
			\node[location, initial] at (0,0) (l1) {$\styleloc{\loc_1}$};
	
			\node[location, right of=l1] (l2) {$\styleloc{\loc_2}$};
			
			\node[location, right of=l2] (l3) {$\styleloc{\loc_3}$};

			\path (l1) edge[] node[above]{$ \styleclock{x_1} = \styleparam{\param}$} node[below] {$ \styleclock{x_1} := 0$} (l2);

			\path (l2) edge[] node[above]{$ \styleclock{x_1} = \styleparam{\param} \land \styleclock{x_2} = 1$} node[below] {$ \styleclock{x_1} := 0$} (l3);
		\end{tikzpicture}

		\bigskip

		\begin{tikzpicture}[scale=1, node distance=3.5cm, auto, ->, >=stealth']
		
		\footnotesize
			\node[initial,pzgState] (s0){$\styleloc{\loc_1}$
				\nodepart{two}\shortstack[l]{$\styleclock{\clock_2}\geq0$\\
						$\styleparam{\param}\geq0$\\
						$\styleclock{\clock_1} = \styleclock{\clock_2}$}};
			\node[pzgState, right of=s0] (s1){$\styleloc{\loc_2}$
				\nodepart{two}\shortstack[l]{%
						$\styleclock{\clock_2}\geq\styleparam{\param}$\\
						$\styleparam{\param}\geq0$\\
						$\styleclock{\clock_2} = \styleclock{\clock_1}
							+ \styleparam{\param}$}};
			\node[pzgState, right of=s1] (s2){$\styleloc{\loc_3}$
				\nodepart{two}\shortstack[l]{$\styleclock{\clock_2}\geq1$\\
						$2\styleparam{\param} = 1$\\
						$\styleclock{\clock_2} = \styleclock{\clock_1} + 1$}};
			\draw [->] (s0) -- (s1);
			\draw [->] (s1) -- (s2);
		\end{tikzpicture}
		
		\caption{A PTA which is not an IP-PTA, and its Parametric Zone Graph}
		\label{fig:IPPTA:no}
	\end{subfigure}
	\\
	\begin{subfigure}[b]{\textwidth}
		\centering
		
		\begin{tikzpicture}[scale=1, auto, ->, >=stealth']
		
		\footnotesize
	
			\node[location, initial] at (0,0) (l1) {$\styleloc{\loc_1}$};
			\node [invariant,below] at (l1.south) {$ \styleclock{x} \leq 1$};
	
			\node[location,final] at (3,0) (l2) {$\styleloc{\loc_2}$};
			
			\node[location, final] at (0,1.5) (l3) {$\styleloc{\loc_3}$};

			\path (l2) edge[loop above] node[above]{\begin{tabular}{@{} c @{\ } c@{} }
			& $ \styleclock{x} = 1$\\
			& $\styleact{c}$\\
			& $\styleclock{x}:=0$\\
			\end{tabular}} (l2);

			\path (l1) edge[] node[above]{\begin{tabular}{@{} c @{\ } c@{} }
			& $ \styleclock{x} = 1$\\
			$\land$ & $ \styleclock{x} \leq \styleparam{\param_2}$\\
			& $\styleact{b}$\\
			& $\styleclock{x}:=0$\\
			\end{tabular}} (l2);

			\path (l1) edge[] node[left]{\begin{tabular}{@{} c @{\ } c@{} }
			& $ \styleclock{x} = \styleparam{\param_1}$\\
			& $\styleact{a}$\\
			\end{tabular}} (l3);

		\footnotesize
			\node[initial,pzgState] at (5.5,0) (s0){$\styleloc{\loc_1}$
				\nodepart{two}\shortstack[l]{%
						$0\leq\styleclock{\clock}\leq1$\\
						$\styleparam{\param_1}\geq0$\\
						$\styleparam{\param_2}\geq0$}};
			\node[pzgState] at (9,1) (s1){$\styleloc{\loc_3}$
				\nodepart{two}\shortstack[l]{%
						$0\leq\styleparam{\param_1}\leq1$\\
						$\styleparam{\param_2}\geq0$\\
						$\styleclock{\clock}\geq\styleparam{\param_1}$}};
			\node[pzgState] at (9,-1) (s2){$\styleloc{\loc_2}$
				\nodepart{two}\shortstack[l]{%
						$\styleclock{\clock}\geq0$\\
						$\styleparam{\param_1}\geq0$\\
						$\styleparam{\param_2}\geq1$}};
			\draw [->] (s0) -- node[above] {$\styleact{a}$} (s1);
			\draw [->] (s0) -- node[above] {$\styleact{b}$} (s2);
			\path [->] (s2) edge[loop right,min distance=5mm]
				node[right] {$\styleact{c}$} (s2);
		\end{tikzpicture}
	
		\caption{A PTA which is an IP-PTA, and its Parametric Zone Graph}
		\label{fig:exULTL}
	\end{subfigure}

	\caption{Examples of PTA}
\end{figure}
\begin{example}
	Consider the PTA in \cref{fig:IPPTA:no}, containing two clocks $\clocki{1}$ and~$\clocki{2}$, and one parameter $\styleparam{\param}$.
	This PTA is not an IP-PTA. Indeed, as can be seen on its parametric zone graph, the (unique) symbolic state with location~$
	\styleloc{\loc_3}$ contains only $\styleparam{\param}=\frac{1}{2}$, and this symbolic state therefore contains no integer point.
	
	In contrast, the PTA in \cref{fig:exULTL} is an IP-PTA: each zone in its parametric
	zone graph contains an integer valuation of all parameters.
	The coffee machine in \cref{fig:coffee} (which has an infinite parametric zone graph) is also an IP-PTA. 
\end{example}

In~\cite{ALR16ICFEM}, we studied the expressiveness of IP-PTAs:
while the class of IP-PTAs is incomparable with the class of L/U-PTAs, any \emph{non-strict} L/U-PTA, \ie{} with only non-strict inequalities, is an IP-PTA.

Concerning decidability, the only non-trivial general class with a decidability result for \EF{}-emptiness is L/U-PTAs~\cite{HRSV02}.
We extended this class, by proving that \EF{}-emptiness is decidable for bounded IP-PTAs~\cite{ALR16ICFEM}.
However, other studied problems turned out to be undecidable.

\subsubsection{Summary}

\newcommand\crefabbr[1]{%
\begingroup
	\crefname{theorem}{\text{Th.}}{\text{Th.}}
	\crefname{corollary}{\text{Cor.}}{\text{Cor.}}
	\cref{#1}
\endgroup%
}

\begin{table*}[tb!]
	\centering
	\setlength{\tabcolsep}{3pt} %
	\begin{tabular}{@{} | c | c | c | c | c | c | c | c | c |}
		\hline
		\cellHeader{Class}
			& \cellHeader{U-PTAs}
			& \multicolumn{2}{c |}{\cellHeader{bL/U-PTAs}}
			& \cellHeader{L/U-PTAs}
			& \cellHeader{bIP-PTAs}
			& \cellHeader{IP-PTAs}
			& \cellHeader{bPTAs}
			& \cellHeader{PTAs}
		\\
		\rowHeader{} & & \cellHeader{closed} & \cellHeader{open} & & & & & \\

		\hline
		\EF{}
			& \colCellDec{}\cite{HRSV02}
			& \colCellDecNous{}\cite{ALR16ICFEM} %
			& \cellOpen{}
			& \colCellDec{}\cite{HRSV02}
			& \colCellDecNous{}\cite{ALR16ICFEM}
			& \colCellUndecNous{}\cite{ALR16ICFEM}
			& \colCellUndec{}\cite{Miller00}
			& \colCellUndec{}\cite{AHV93}
		\\

		\hline
		\AF{}
			& \cellOpen{}
			& \multicolumn{2}{ c |}{\colCellUndecNous{}\cite{ALR16ICFEM}} %
			& \colCellUndec{}\cite{JLR15}
			& \colCellUndecNous{}\cite{ALR16ICFEM}
			& \colCellUndecNous{}\cite{ALR16ICFEM}
			& \colCellUndecNous{}\cite{ALR16ICFEM}
			& \colCellUndec{}\cite{JLR15}
		\\

		\hline
		\EG{}
			& \cellOpen{}
			& \colCellDecNous{}\cite{ALime17}
			& \colCellUndecNous{}\cite{ALime17}
			& \colCellUndecNous{}\cite{ALime17}
			& \multicolumn{2}{ c |}{\cellOpen{}}
			& \multicolumn{2}{ c |}{\colCellUndecNous{}\cite{ALime17}}
		\\

		\hline
		\AG{}
			& \colCellDecNous{}\cite{ALime17} %
			& \colCellDecNous{}\cite{ALR16ICFEM} %
			& \cellOpen{}
			& \colCellDecNous{}\cite{ALime17}
			& \multicolumn{4}{ c |}{\colCellUndecNous{}\cite{ALR16ICFEM}}
		\\

		\hline
		TCTL
			& \multicolumn{1}{c |}{\colCellUndecNous{\cite{ALR18FORMATS}}}
			& \multicolumn{2}{ c |}{\colCellUndecNous{\cite{ALR16ICFEM}}}
			& \multicolumn{1}{c |}{\colCellUndec{\cite{JLR15}}}
			& \multicolumn{2}{ c |}{\colCellUndecNous{\cite{ALR16ICFEM}}}
			& \multicolumn{1}{ c |}{\colCellUndec\cite{Miller00}}
			& \multicolumn{1}{ c |}{\colCellUndec{}\cite{AHV93}}
		\\
		
		\hline
		\hline
		EC
			& \colCellDecNous{}\cite{ALime17}
			& \colCellDecNous{}\cite{ALime17}
			& \cellOpen{}
			& \colCellDecNous{}\cite{ALime17}
			& \multicolumn{2}{ c |}{\cellOpen{}}
			& \multicolumn{2}{ c |}{\colCellUndecNous{}\cite{ALime17}}
		\\

		\hline
		ED
			& \cellOpen{}
			& \multicolumn{3}{ c |}{\colCellUndecNous{}\cite{ALime17}}
			& \multicolumn{2}{ c |}{\cellOpen{}}
			& \colCellUndecNous{}\cite{ALime17}
			& \colCellUndecNous{}\cite{Andre16}
		\\

		\hline
		\hline
		LgP
			& \cellOpen{}
			& \multicolumn{3}{ c |}{\colCellUndecNous{}\cite{AM15}}
			& \multicolumn{2}{ c |}{\cellOpen{}}
			& \multicolumn{2}{ c |}{\colCellUndecNous{}\cite{AM15}}
		\\

		\hline
		TrP
			& \cellOpen{}
			& \multicolumn{3}{ c |}{\colCellUndecNous{}\cite{AM15}}
			& \multicolumn{2}{ c |}{\cellOpen{}}
			& \multicolumn{2}{ c |}{\colCellUndecNous{}\cite{AM15}}
		\\

		\hline
	\end{tabular}

	\caption{Decidability of the emptiness problems for PTAs and subclasses}
    \label{table:summary:decidability}
\end{table*}

\cref{table:summary:decidability} summarises the decidability results.
It gives from left to right the (un)decidability for U-PTAs, bounded L/U-PTAs (with
either closed or open bounds), L/U-PTAs, bounded IP-PTAs, IP-PTAs, 
bounded PTAs, and PTAs.
We review the emptiness of TCTL subformulas (\EF, \AF, \EG, \AG{}), full TCTL, cycle-existence,
deadlock-existence and language- and trace-preservation.
Decidability is given in green, whereas undecidability is given in italic red.
Our contributions are emphasized in bold using a plain background, whereas existing results are depicted using a light background.
When several papers in the literature proved the same result, we only give the earliest result, and not necessarily the best (in terms of number 
of clocks and parameters, or complexity).

\paragraph{Perspective: open subclasses}
L-PTAs and U-PTAs~\cite{BlT09} are very open classes, in the sense that the only known decidability results come from the larger class of PTAs, 
and no undecidability result was known---with the exception of our recent result concerning TCTL-emptiness~\cite{ALR18FORMATS}.
To summarize, the \EG{}-emptiness, \AG{}-emptiness and \AF{}-emptiness problems, as well as the language- and trace-preservation problems, 
are all undecidable for (general) L/U-PTAs, but remain open for L-PTAs and U-PTAs.
Similarly, the \EF-synthesis problem (shown intractable for L/U-PTAs in~\cite{JLR15} despite the decidability of the \EF-emptiness problem) 
remains open for rational-valued L- and U-PTAs, and would significantly increase the interest of these subclasses if it was shown to be computable.

\section{Parametric Interval Markov Chains}
\label{sec:PIMC}

Parametric probabilities are useful to capture imprecisions, robustness and dimensioning
issues. Hence, in this section we consider Parametric Interval Markov Chains (PIMC).

\subsection{Introduction to Parametric Markov Chains}
\label{sec:PIMC:intro}

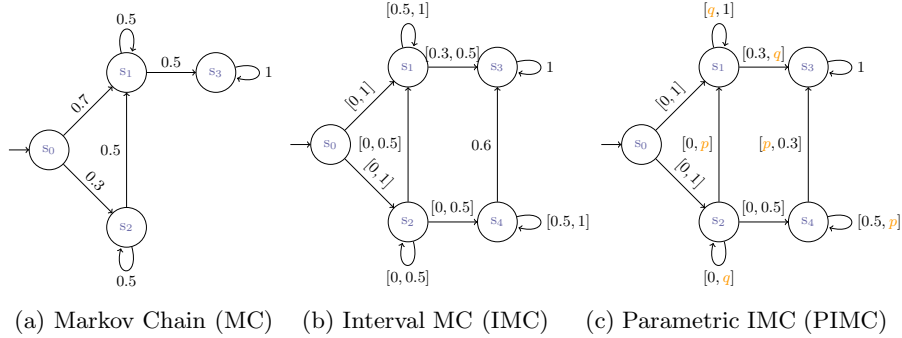
\begin{figure}[!!htb]
	\begin{subfigure}[b]{.33\textwidth}
	\scalebox{0.65}{
    \begin{tikzpicture}
        \node[state,initial] (s0) {$\styleloc{s_0}$};
        \node[state, above right=of s0] (s1) {$\styleloc{s_1}$};
        \node[state, below right=of s0] (s2) {$\styleloc{s_2}$};
        \node[state, right=of s1] (s3) {$\styleloc{s_3}$};

        \draw[->] (s0) edge node[above, sloped] {$0.7$} (s1);
        \draw[->] (s0) edge[swap] node[above,sloped] {$0.3$} (s2);
        \draw[->] (s1) edge[loop above] node {$0.5$} ();
        \draw[->] (s1) edge node[above,sloped] {$0.5$} (s3);
        \draw[->] (s2) edge[loop below] node {$0.5$} ();
        \draw[->] (s2) edge[swap] node[left] {$0.5$} (s1);
        \draw[->] (s3) edge[loop right] node {$1$} ();
    \end{tikzpicture}
    }
	\caption{Markov Chain (MC)}
	\label{fig:MC}
	\end{subfigure}
\hspace{-0.5cm}
	\begin{subfigure}[b]{.33\textwidth}
	\scalebox{0.65}{
    \begin{tikzpicture}
        \node[state,initial] (s0) {$\styleloc{s_0}$};
        \node[state, above right=of s0] (s1) {$\styleloc{s_1}$};
        \node[state, below right=of s0] (s2) {$\styleloc{s_2}$};
        \node[state, right=of s1] (s3) {$\styleloc{s_3}$};
        \node[state, right=of s2] (s4) {$\styleloc{s_4}$};

        \draw[->] (s0) edge node[above,sloped] {$[0,1]$} (s1);
        \draw[->] (s0) edge[swap] node[above,sloped] {$[0,1]$} (s2);
        \draw[->] (s1) edge[loop above] node {$[0.5,1]$} ();
        \draw[->] (s1) edge node[above,sloped] {$[0.3,0.5]$} (s3);
        \draw[->] (s2) edge[loop below] node {$[0,0.5]$} ();
        \draw[->] (s2) edge[swap] node[left] {$[0,0.5]$} (s1);
        \draw[->] (s2) edge node[above,sloped] {$[0,0.5]$} (s4);
        \draw[->] (s4) edge[swap] node[left] {$0.6$} (s3);
        \draw[->] (s3) edge[loop right] node {$1$} ();
        \draw[->] (s4) edge[loop right] node {$[0.5,1]$} ();
    \end{tikzpicture}
    }
	\caption{Interval MC (IMC)}
	\label{fig:IMC}
	\end{subfigure}
	\begin{subfigure}[b]{.33\textwidth}
	\scalebox{0.65}{
    \begin{tikzpicture}
        \node[state,initial] (s0) {$\styleloc{s_0}$};
        \node[state, above right=of s0] (s1) {$\styleloc{s_1}$};
        \node[state, below right=of s0] (s2) {$\styleloc{s_2}$};
        \node[state, right=of s1] (s3) {$\styleloc{s_3}$};
        \node[state, right=of s2] (s4) {$\styleloc{s_4}$};

        \draw[->] (s0) edge node[above,sloped] {$[0,1]$} (s1);
        \draw[->] (s0) edge[swap] node[above,sloped] {$[0,1]$} (s2);
        \draw[->] (s1) edge[loop above] node {$[\styleparam{q},1]$} ();
        \draw[->] (s1) edge node[above,sloped] {$[0.3,\styleparam{q}]$} (s3);
        \draw[->] (s2) edge[loop below] node {$[0,\styleparam{q}]$} ();
        \draw[->] (s2) edge[swap] node[left] {$[0,\styleparam{p}]$} (s1);
        \draw[->] (s2) edge node[above,sloped] {$[0,0.5]$} (s4);
        \draw[->] (s4) edge[swap] node[left] {$[\styleparam{p},0.3]$} (s3);
        \draw[->] (s3) edge[loop right] node {$1$} ();
        \draw[->] (s4) edge[loop right] node {$[0.5,\styleparam{p}]$} ();
    \end{tikzpicture}
    }
	\caption{Parametric IMC (PIMC)}
	\label{fig:PIMC}
	\end{subfigure}
\caption{Markov Chains and their extensions\label{fig:ex:MC}}
\end{figure}

\cref{fig:ex:MC} contains an example of the different flavours of Markov Chains we
are addressing in this section. \cref{fig:MC} is a \emph{Markov chain} (MC). As in
an automaton,
there are states and transitions between them, but these are labelled by probabilities
for the transition to occur. For example, from state $\styleloc{s_0}$, there is a probability
of $0.7$ to go to state $\styleloc{s_1}$ and of $0.3$ to go to state $\styleloc{s_2}$. Therefore, the
sum of all probabilities labelling transitions exiting a state must be $1$.

When probabilities are not known in advance, it might still be possible to know an
interval
to which they belong. Hence, we introduce \emph{Interval Markov Chains} (IMC), as
pictured in \cref{fig:IMC}. The transitions are then no more labelled by a fixed probability
but an interval meaning that the probability should be between the lower and the upper
bound of the interval. For example, the probability to move from state $\styleloc{s_1}$ to state
$\styleloc{s_3}$ is between $0.3$ and $0.5$. The MC in \cref{fig:MC} can be seen as an implementation of the IMC in
\cref{fig:IMC} which stands as a specification. Notice that state $\styleloc{s_4}$ does not
appear in \cref{fig:MC}, which is equivalent to having a transition from state
$\styleloc{s_2}$ to state $\styleloc{s_4}$ with a probability $0$. 
Once a probability is chosen in an interval, it imposes
constraints on the other probabilities outgoing the same state since they must add
up to $1$. An IMC is said to be \emph{consistent} if it admits at least one implementation.

When the upper or lower bounds are unknown, it is convenient to use parameters. \cref
{fig:PIMC} shows a \emph{Parametric Interval Markov Chain} (PIMC). As compared with
the IMC of \cref{fig:IMC}, some of the bounds are replaced with parameters $\styleparam{p}$
and $\styleparam{q}$. Notice that the same parameter occurs at several places in the
PIMC, therefore imposing constraints on the interval. For example, from state $\styleloc{s_1}$,
it is possible to stay in this state with a probability between $\styleparam{q}$
and $1$, or to move to state $\styleloc{s_3}$ with a probability between $0.3$ and the same
$\styleparam{q}$. When assigning a valuation to all parameters in a PIMC, we obtain an IMC.

\subsection{Markov Chains definitions}

We now formally define the different Markov Chain models. These definitions are detailed
in \cite{BDDLLP-VMCAI16}.

\begin{definition}[Markov Chain]
\label{def:mc}  
A \emph{Markov Chain} is a tuple $\mathcal{M} = (S, s_0, M, A,\linebreak[4] V)$,
where $S$
is a finite set of states containing the initial state $s_0$, $A$ is a
set of atomic propositions, $V : S \rightarrow 2^A$ is a labelling
function, and $M : S \times S \rightarrow [0,1]$ is a probabilistic
transition function such that $\forall s \in S, \sum_{t \in S} M(s,t)
= 1$.
\end{definition}

We now introduce the notation of parameters, and interval ranges that will be used
throughout this section. A parameter
$p\in P$ is a variable ranging through the interval $[0,1]$. A
valuation for $P$ is a function $\psi : P \rightarrow [0,1]$ that
associates values with each parameter in $P$. We write
$\Int_{[0,1]}(P)$ for the set of all closed parametric intervals of the form
$[x,y]$ with $x,y \in [0,1] \cup P$. When $P = \emptyset$, we write
$\Int_{[0,1]} = \Int_{[0,1]}(\emptyset)$ to denote closed intervals
with real-valued endpoints. Given an interval $I$ of the form $I =
[a,b]$, $\low(I)$ and $\up(I)$ respectively denote the lower and upper
endpoints of $I$, i.e. $a$ and $b$. Given an interval $I = [a,b] \in
\Int_{[0,1]}$, we say that $I$ is well-formed whenever $a \le b$.

The definition of Interval Markov Chains is adapted
from~\cite{DBLP:journals/jlp/DelahayeLLPW12}. 

\begin{definition}[Interval Markov Chain~\cite{DBLP:journals/jlp/DelahayeLLPW12}]
An \emph{Interval Markov Chain} is a tuple $\mathcal{I} = (S, s_0,
\phi, A, V)$, where $S$, $s_0$, $A$ and $V$ are as for MCs,
and $\phi : S \times S \rightarrow \Int_{[0,1]}$ is a transition
constraint that associates with each potential transition an interval
of probabilities.
\end{definition}

The following definition recalls the notion of satisfaction introduced
in~\cite{DBLP:journals/jlp/DelahayeLLPW12}. Satisfaction (also called
implementation in some cases) allows to characterise the set of MCs
represented by a given IMC specification. Satisfaction
abstracts from the syntactic structure of transitions in IMCs: a
single transition in the implementation MC can contribute to
satisfaction of more than one transition in the specification IMC, by
distributing its probability mass against several transitions.
Similarly many MC transitions can contribute to the satisfaction of
just one specification transition. This crucial notion is embedded in
the so-called {\em correspondence function} $\delta$ introduced
below. Informally, such a function is given for all pairs of states
$(t,s)$ in the satisfaction relation, and associates with each successor
state $t'$ of $t$ -- in the implementation MC -- a distribution over
potential successor states $s'$ of $s$ -- in the specification IMC --
specifying how the transition $t \rightarrow t'$ contributes to
the transition $s \rightarrow s'$.

\begin{definition}[Satisfaction Relation~\cite{DBLP:journals/jlp/DelahayeLLPW12}]
\label{def:satisfaction}
Let $\mathcal{I} = (S, s_0, \phi, A, V^I)$ be an IMC and $\mathcal{M} =
(T, t_0, M, A, V^M)$ be a MC. A relation $\rel \subseteq T
\times S$ is a {\em satisfaction relation} if whenever $t \rel s$, 
\begin{enumerate}
\item the labels of $s$ and $t$ agree: $V^M(t) = V^I(s)$,
\item there exists a {\em correspondence} function $\delta : T \rightarrow (S \rightarrow
  [0,1])$ such that \begin{enumerate} \item for all $t' \in T$ such
  that $M(t,t') >0$, $\delta(t')$ is a distribution on $S$, \item for
  all $s' \in S$, we have $(\sum_{t' \in T}
  M(t,t') \cdot \delta(t')(s')) \in \phi(s,s')$, and \item for all
  $t' \in T$ and $s' \in S$, if $\delta(t')(s') > 0$, then
  $(t',s') \in \rel$.  \end{enumerate} 

We say that state $t \in T$ satisfies state $s \in S$ (written
  $t \models s$) iff there exists a (minimal) satisfaction relation
  containing $(t,s)$ and that $\mathcal{M}$ satisfies $\mathcal{I}$
  (written $\mathcal{M} \models \mathcal{I}$) iff $t_0 \models s_0$.
\end{enumerate}
\end{definition}

The set of MCs satisfying a given IMC $\mathcal{I}$ is written
$\impl{\mathcal{I}}$. Formally, $\impl{\mathcal{I}}
= \{\mathcal{M} \st \mathcal{M} \models \mathcal{I}\}$.

\begin{definition}
An IMC $\mathcal{I}$ is {\em consistent} iff $\impl{\mathcal{I}} \ne \emptyset$.
\end{definition}

Although
the satisfaction relation abstracts from the syntactic structure of
transitions, we recall the following result from \cite{pIMC-syncop},
that states that whenever a given IMC is consistent, it admits at
least one implementation that strictly respects its structure.

\begin{theorem}[\cite{pIMC-syncop}]
\label{thm:structure}
An IMC $\mathcal{I} = (S, s_0, \phi, A, V)$ is consistent iff it
admits an implementation of the form $\mathcal{M} = (S, s_0, M, A, V)$
where, for all reachable states $s$ in $\mathcal{M}$, it holds that
$M(s,s') \in \phi(s,s')$ for all $s'$.
\end{theorem}

In the following, we say that state $s$ is consistent in the IMC
$\mathcal{I} = (S,s_0,\phi,A,V)$ if there exists an implementation
$\mathcal{M} = (S,s_0,M,A,V)$ of $\mathcal{I}$ in which state $s$ is
reachable with a non-zero probability.

We now recall to the notion of Parametric Interval Markov Chain, previously introduced in~\cite{pIMC-syncop}.

\begin{definition}[Parametric Interval Markov Chain]
A \emph{Parametric Interval Markov Chain} is a tuple $\mathcal{I}^P =
(S, s_0, \phi_P, A, V, P)$, where $S$, $s_0$, $A$ and $V$ are as for IMCs,
$P$ is a set of variables (parameters) ranging over $[0,1]$ and
$\phi_P : S\times S \rightarrow \Int_{[0,1]}(P)$ associates with each
potential transition a (parametric) interval.
\end{definition}

Given a pIMC $\mathcal{I}^P=(S,s_0,\phi_P,A,V,P)$ and a parameter
valuation $\psi:P\rightarrow [0,1]$, we write $\psi(\mathcal{I}^P)$
for the IMC obtained by replacing $\phi_P$ by the function
$\phi:S\times S\rightarrow\Int_{[0,1]}$ defined by $\forall s,s'\in
S, \phi(s,s') = \psi(\phi_P(s,s'))$. The IMC $\psi(\mathcal{I}^P)$
is called an \emph{instance} of pIMC $\mathcal{I}^P$.

Finally, we say that a MC $\mathcal{M} = (T, t_0, M, A, V^M)$ {\em
implements} pIMC $\mathcal{I}^P$, written
$\mathcal{M} \models \mathcal{I}^P$, iff there exists an instance
$\mathcal{I}$ of $\mathcal{I}^P$ such that
$\mathcal{M} \models \mathcal{I}$. We write $\impl{\mathcal{I}^P}$ for
the set of MCs implementing $\mathcal{I}^P$ and say that a pIMC is
{\em consistent} iff its set of implementations is not empty.

\subsection{Consistency of PIMCs}
\label{sec:PIMC:consistent}

When considering IMCs, one question of interest is to decide whether
it is consistent without computing its set of implementations. This
problem has been addressed in~\cite{DBLP:journals/jlp/DelahayeLLPW12,pIMC-syncop},
yielding polynomial decision algorithms and procedures that produce
one implementation when the IMC is consistent. The same question holds
for pIMCs, although in a slightly different setting.
\cite{pIMC-syncop} proposed a polynomial algorithm for
deciding whether a given pIMC is consistent, in the sense that it
admits at least one parameter valuation for which the resulting IMC is
consistent. 

In order to decide whether a given IMC is consistent, we need to
address the set of potential successors of a given state
$s$. Let $\suc(s)$ be the set of states that can be reached from $s$ with a probability interval not
reduced to $[0,0]$: $\suc(s) = \{s' \in S \st \phi_P(s,s')\ne [0,0]\}$.

We now introduce the notion of $n$-consistency in the IMC setting and
then adapt this notion to pIMCs. In practice, $n$-consistency is defined by induction over
the structure of $\mathcal{I}$.

\begin{definition}[$n$-consistency]
    Let $\mathcal{I}=(S,s_0,\phi,A,V)$ be an IMC and let
    $\D:S\rightarrow\Dist(S)$ be a function that assigns a
    distribution on $S$ to each state of $\mathcal{I}$. State $s\in S$ is
    $(n,\D)$-consistent iff for all $s'\in S$,
    $\D(s)(s')\in\phi(s,s')$, and, for $n>0$, $\D(s)(s')>0$ implies
    $s'$ is $(n-1,\D)$-consistent.

    We say that $s$ is $n$-consistent if there exists
    $\D:S\rightarrow\Dist(S)$ such that $s$ is $(n,\D)$-consistent.
\label{def:n-consistency}
\end{definition}

\cref{def:n-consistency} is thus equivalent to the
following intuitive inductive definition: a state $s$ is
$n$-consistent iff there exists a distribution $\rho$ satisfying all
of its outgoing probability intervals and such that for all $s' \in
S$, $\rho(s')>0$ implies that $s'$ is $(n-1)$-consistent.

\begin{theorem}
    Given an IMC $\mathcal{I} = (S,s_0,\phi,A,V)$, $\mathcal{I}$ is consistent iff $s_0$ is $|S|$-consistent.
    \label{thm:consS}
\end{theorem}

\begin{example}
Let us consider the example in \cref{fig:IMC}. All states but $\styleloc{s_4}$ are
$0$-consistent.
Indeed it is possible to find probabilities within the exiting intervals that add
up to $1$. It is not the case for state $\styleloc{s_4}$, which has a probability of $0.6$ to which
we should add one that is at least $0.5$, so the sum is at least $1.1$.
Then, for state $\styleloc{s_2}$ to be $1$-consistent, it must not have $\styleloc{s_4}$ has a successor. This
is possible by choosing $0$ as the probability to go from $\styleloc{s_2}$ to $\styleloc{s_4}$. In the remaining
two intervals, choosing probability $0.5$ leads to a sum of $1$. Therefore, state
$\styleloc{s_2}$ is $1$-consistent. One can check that all states but $\styleloc{s_4}$ are $n$-consistent, for
all $n$.
\end{example}

For the problem of consistency of pIMCs, the aim is not only to decide whether a given
pIMC is consistent, but also to synthesise all parameter valuations that
ensure consistency of the resulting IMC. For this purpose, we adapt
the notion of $n$-consistency defined above to pIMCs.

We first define the \emph{local consistency} of a state \wrt{} some subset $S'$ of
its successors: the sum of upper bounds should be greater than $1$, the sum of lower
bounds smaller than $1$, and all successors in $S'$ have a valid interval.

\centerline{\scalebox{.94}{\begin{minipage}{\linewidth}
\begin{multline*}
LC(s,S') = \left [ \sum_{s' \in S'} \up(\phi_P(s,s')) \ge 1 \right ]
\cap \left [ \sum_{s' \in S'} \low(\phi_P(s,s')) \le 1 \right ] \\
 \cap \left [ \bigcap_{s' \in S'} \low(\phi_P(s,s')) \le
  \up(\phi_P(s,s')) \right ]
\end{multline*}
\end{minipage}}}

Let us start by fixing a set of
states $X$ that we want to avoid and then compute the set of
valuations $\cons_n^{X}(s)$ that ensure $n$-consistency of $s$ through a
distribution $\rho$ that avoids states from $X$. Formally,
$\cons_n^X(s)$ is defined as: let
$\cons^X_0(s) = LC(s,\suc(s)\setminus X)\cap \left [ \bigcap_{s'\in
X} \low(\phi_P(s,s')) = 0 \right ]$ and for $n \ge 1$,

\vspace{-.3cm}

\centerline{\scalebox{.94}{\begin{minipage}{\linewidth}
\begin{multline*}
\cons^X_n(s) = 
     \left [ \bigcap_{s' \in \suc(s) \setminus X} \cons_{n-1}(s') \right ] 
\cap \left [ LC(s, \suc(s)\setminus X) \right ] \\
\cap \left [ \bigcap_{s'\in X} \low(\phi_P(s,s')) = 0 \right ]
\end{multline*}
\end{minipage}}}

The set of valuations ensuring $n$-consistency is then the union, for
all potential choices of $X$, of $\cons_n^X(s)$. We need to choose $X$ as a
subset of the set $Z(s)$ of states which can be avoided, by transitions that have
$0$ or a parameter as lower bound. Therefore, we define
$\cons_n(s)=\bigcup_{X \subseteq Z(s)} \cons^X_n(s)$.

\begin{theorem}
Given a pIMC $\mathcal{I}^P = (S, s_0, \phi_P, A, V, P)$ and a
parameter valuation $\psi:P\rightarrow [0,1]$, we have
$\psi \in \cons_{|S|}(s_0)$ iff the IMC $\psi(\mathcal{I}^P)$ is consistent.
\label{thm:consistency}
\end{theorem}

\begin{example}
It is easily shown that the consistency of the PIMC in \cref{fig:PIMC} is $[(\styleparam{q}\leq0.7)\cap
(\styleparam{q}\geq0.3)]\cup(\styleparam{q}=1)$.
\end{example}

This section has shown the crucial property of consistency in both parametric and
non-parametric interval Markov chains. It thus sets the necessary elements before
model checking such probabilistic models.

\subsection{Further reading}
\label{sec:further}

The definitions and properties stated in this section are detailed in \cite{BDDLLP-VMCAI16}.
They were revised in~\cite{LP-JP-FORTE-18}, where
both inductive and co-inductive
definitions of consistency are given, implemented with forward and backward algorithms.
\cite{BDDLLP-VMCAI16} also addresses some properties: consistent avoidability, existential and
universal consistent reachability.

\section{Parametric Petri Nets}
\label{sec:ppn}

\newcommand{\Net}{\mathcal{N}}
\newcommand{\SetPar}{\mathbb{P}}
\newcommand{\natN}{\mathbb{N}}
\newcommand{\RS}{\mathcal{RS}}
\newcommand{\val}{\ensuremath{\vec{v}}}
\newcommand{\Places}{\ensuremath{\mathsf{P}}}
\newcommand{\place}{\ensuremath{p}}
\newcommand{\Trans}{\ensuremath{\mathsf{T}}}
\newcommand{\trans}{\ensuremath{t}}
\newcommand{\Pre}{\ensuremath{\mathsf{Pre}}}
\newcommand{\Post}{\ensuremath{\mathsf{Post}}}
\newcommand{\marking}{\ensuremath{m}}
\newcommand{\fire}[1]{\overset{#1}{\rightarrow}}
\newcommand{\EXPSPACE}{{\sc ExpSpace}}

\newcommand{\cmark}{\ding{51}}%
\newcommand{\xmark}{\ding{53}}%

We now consider a parametric extension of Petri nets in which the weights of the
arcs can be parameters.
This extension was mainly studied in the Ph.D.\ thesis of Nicolas David~\cite{David17} 
and many of those results can also be found in~\cite{DJLR15,DJLR17}.

\begin{example}
In order to illustrate the usefulness of this parameterised formalism, we consider the example, 
taken from~\cite{David17}, of a financial loan. It is modelled in~\cref{fig:financial}.
		\begin{figure}[h!] \centering
            \scalebox{0.7}{%
	\begin{tikzpicture}[node distance=1.3cm,>=stealth',bend angle=45,auto]

        \begin{scope}
    \node [place, tokens=1] (p1) [xshift=-26mm,label={[black]below:\footnotesize{$
    \styleloc{initialisation}$}}] {$$};
    \node [place] (p2) [xshift=17mm,yshift=0mm,label={[black]below:\footnotesize{$\styleloc{funds}$}}] {$a$};
    \node [place] (p3) [xshift=17mm,yshift=13mm,label={[black]above:\footnotesize{$\styleloc{months}$}}] {$$};
    \node [place] (p4) [xshift=30mm,yshift=-13mm,label={[black]below:\footnotesize{$\styleloc{lock_1}$}}] {$$};
    \node [place] (p5) [xshift=65mm,yshift=0mm,label={[black]below:\footnotesize{$\styleloc{total}$}}] {$$};
    \node [place] (p6) [xshift=91mm,yshift=0mm,label={[black]below:\footnotesize{$\styleloc{loanOk}$}}] {$$};
    \node [place] (p7) [xshift=91mm,yshift=13mm,label={[black]above:\footnotesize{$\styleloc{interestOk}$}}] {$$};
    \node [place] (p8) [xshift=120mm,yshift=0mm,label={[black]above:\footnotesize{$\styleloc{loanFinished}$}}] {$$};
    \node [place] (plock3) [xshift=78mm,yshift=-13mm,label={[black]below:\footnotesize{$\styleloc{lock_2}$}}] {$$};
     \node [place] (plock4) [xshift=66mm,yshift=26mm,label={[black]below:\footnotesize{$\styleloc{lock_3}$}}] {$$};
    
    \node [transition] (t1) [left of=p2, label={[black]above left:\footnotesize{$\styleact{grantLoan}$}}] {}
      edge [pre] (p1)
      edge [post] node{$\styleparam{c}$}(p2)
      edge [post, bend left] node{$\styleparam{d}$}(p3)
      edge [post, out=-90, in = 180] node{$$}(p4);

    \node [transition] (t2) [right of=p2, label={[black]above right:\footnotesize{$\styleact{reimburse}$}}] {}
      edge [pre] (p2)
      edge [pre,bend right] (p3)
      edge [pre, bend left] (p4)
      edge [post,bend right] node[label={above:$\styleparam{b}$}]{}(p2)
      edge [post,bend right] (p4)
      edge [post] node[label={ [shift={(-0.3,-0.7)}]$\styleparam{e}$}]{}(p5) ;
      
    \node [transition] (t2bis) [right of=p4, xshift=9mm,label={[black]below:\footnotesize{$\styleact{endLoan}$}}] {}
    	edge [pre] (p4)
	edge [post] (plock3)
	edge [post, out=90,in=180] (plock4);
	
    \node [transition] (t3) [right of=p5, label={[shift={(-0.1,-0.1)},black]\footnotesize{$\styleact{getAmount}$}}] {}
      edge [pre] node{$\styleparam{c}$} (p5)
      edge [pre] (plock3)
      edge [post] (p6);
    \node [transition] (t4) [right of=p5, above of =p5, label={[shift={(0,-0.9)},black]\footnotesize{$\styleact{getInterest}$}}] {}
      edge [pre, bend right] node[label={ [shift={(-0.5,-0.6)}]$\styleparam{f}$}]{}
      (p5)
      edge [pre, bend right] (plock4)
      edge [post] (p7);
    \node [transition] (t5) [right of=p6, label={[shift={(0,-0.8)},black]\footnotesize{$\styleact{closeLoan}$}}] {}
      edge [pre, bend right] (p7)
      edge [pre] (p6)
      edge [post] (p8);

  \end{scope}
    \end{tikzpicture}}
			\caption{Modelling a financial loan with parametric Petri nets}
			\label{fig:financial}
		\end{figure}
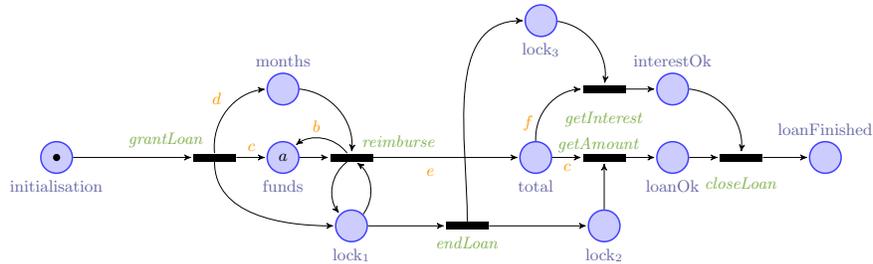
		
		In the general case, a client has a certain amount of money, say $\styleparam{a}$, and is ensured to get income $\styleparam{b}$ every month.
		To finance his project, the client needs to define with the bank the amount
		loaned $\styleparam{c}$, the duration of reimbursement $\styleparam{d}$ (in months) 
		and the amount of each reimbursement $\styleparam{e}$. 
		At the end of the process, the bank expects to get back the initial amount loaned $\styleparam{c}$ plus the interest $\styleparam{f}$.
		The amount of money possessed by the client is depicted in place $\styleloc{funds}$.
		The signature of the contract is symbolised through the firing of transition
		$\styleact{grantLoan}$ which imposes to the bank to loan an amount $\styleparam{c}$ for
		a term of $\styleparam{d}$ months.
		Each month, we can fire $\styleact{reimburse}$: the client receives its own
		income $\styleparam{b}$ that is added to its capital in $\styleloc{funds}$
		and in parallel
		an amount $\styleparam{e}$ is reimbursed to the bank.
		When we consider that the reimbursement is finished, we fire $\styleact{endLoan}$
		that removes the token from $\styleloc{lock_1}$ and allows us to enable the
		transitions
		of the second part of this example.
		We can then check if the bank can get its money back by testing if $\styleloc{loanOk}$
		can be marked and if the bank can get the interest by checking $\styleloc{interestOk}$
		can be marked or both by checking if $\styleloc{loanFinished}$ can be marked.
		\end{example}

        With this example in mind, we now proceed to the corresponding formal definitions.

\begin{definition}[Parametric Petri Net~\cite{DJLR15}]
    A (marked) parametric Petri Net (PPN) is a tuple $\Net=(\Places, \Trans, \SetPar, \Pre, \Post, \marking_0)$ where
    $\Places$ is a finite set of \emph{places},
    $\Trans$ is a finite set of \emph{transitions} such that $\Places\cap\Trans=\emptyset$,
    $\SetPar$ is a finite set of \emph{parameters},
    $\Pre:\Places \times \Trans \rightarrow \natN \cup \SetPar$ is the \emph{backward incidence function}, 
    $\Post:\Places \times \Trans \rightarrow \natN \cup \SetPar$ is the \emph{forward incidence function}, 
    $\marking_{0}\in\natN^{P}$ is the \emph{initial marking}.
\end{definition}
    
    Let $\Net$ be a parametric Petri net. 
    A valuation $\val$ of the parameters of $\Net$ is a mapping from $\SetPar$ to $\natN$.

    We denote by $\val(\Net)$ the \emph{Petri net} obtained by replacing all parameters by the value they are given by $\val$.

    A \emph{marking} of a (non-parametric) Petri net is a mapping from $\Places$ to $\natN$. Markings can be compared component by component: $\marking\geq \marking'$ if $\forall \place\in\Places, \marking(\place)\geq \marking'(\place)$.

    A transition $\trans\in\Trans$ is \emph{enabled} by marking $\marking$ if for all $\place\in\Places$, $\marking(\place)\geq \Pre(\place,\trans)$.
    A transition $\trans$ enabled by marking $\marking$ can be \emph{fired}, leading
    to a new marking $\marking'$ defined by $\forall\place\in\Places, \marking'(\place)
    = \marking(\place) - \Pre(\place,\trans) + \Post(\place,\trans)$. We note $\marking
    \fire{\trans}\marking'$.

    A run in the Petri net is a possibly infinite alternating sequence of markings and transitions $\marking_1\trans_1\marking_2\ldots$ such that $\marking_1=\marking_0$ and for all $i\geq 1$, $\marking_i\fire{\trans_i}\marking_{i+1}$.

\subsection{Problems of interest}

    We extend classic problems defined for Petri nets to the parametric setting.
    These problems are \emph{reachability}, \emph{coverability}, and \emph{(un)boundedness}.

    A marking $\marking$ is \emph{reachable} if there exists a finite run $\marking_1\trans_1\marking_2\ldots\trans_n\marking_
    {n+1}$ such that $\marking_{n+1}=\marking$.

    A marking $\marking$ is \emph{coverable} if there exists a reachable marking $\marking'$ such that $\marking'\geq \marking$.

    A (non-parametric) Petri net is \emph{$k$-bounded} if for all reachable markings
    $\marking$, we have $\forall \place\in\Places, \marking(\place) \leq k$. It is \emph{bounded} if there exists some $k$ such that it is $k$-bounded. If a net is not bounded, we say it is \emph{unbounded} and then for all $B\geq 0$ there exists a place $\place$ and a reachable marking $\marking$ such that $\marking(\place) > B$.

    Similarly a Petri net is \emph{simultaneously $X$ unbounded}~\cite{Demri13}, for
    some subset $X$ of $\Places$, if for all $B\geq 0$, there exists a reachable marking $\marking$ such that for all places $\place\in X$, we have $\marking(\place) > B$.

    We consider two associated \emph{parametric} decision problems: the \emph{existential} and the \emph{universal} problems. In the former we want to decide the existence of a parameter valuation for which some property holds, and in the latter we want to decide if it holds for all the possible parameter valuations. The property in question can be any of those defined above. 
    
    For instance, the existential parametric reachability problem asks, given a target
    marking $\marking$, if there exists a parameter valuation $\val$ 
    such that $\marking$ is reachable in $\val(\Net)$.
    Similarly, the universal coverability problem asks, given a target marking $\marking$, if for all valuations $\val$ of the parameters, $\marking$ is coverable in $\val(\Net)$.

    We finally define \emph{synthesis} problems, in which we want to effectively compute the set of all parameter valuations for which some property holds. Note that if we can effectively compute this set, and check its emptiness or universality, then we can also solve the two decision problems above.

\subsection{Undecidability Results for Parametric Petri Nets}

We start with a few negative results.
\begin{theorem}[\cite{DJLR15}]
    The existential and universal parametric coverability, reachability, and (simultaneous) unboundedness problems are undecidable.
    \label{thm:undec}
\end{theorem}

This theorem is proved by reducing the halting, and counter-boundedness problems for
2-counter machines~\cite{Minsky67} to those parametric problems for parametric Petri
nets. We encode the value of each counter as the number of tokens in a place and we use parametric arcs to test for emptiness of that place (\ie{} counter value $0$).

As a consequence, we need to consider meaningful subclasses for which we might obtain some decidability results. 

\subsection{Subclasses of Parametric Petri Nets}

The basic observation guiding us in defining interesting subclasses of PPNs is that, in the 2-counter machine reduction briefly outlined above, we need both a post arc with a parametric weight $a$ and a pre arc with \emph{the same} parametric weight $a$.

We thus define preT-PPNs, postT-PPNs, and distinctT-PPNs according to whether the parameters are allowed only in pre arcs, in post arcs, or in both but not with the same parameters.

\begin{definition}[preT- and postT-PPNs]
    A preT-PPN (resp. postT-PPN) is a PPN in which the $\Post$ (resp. $\Pre$) function
    has the form $\Places\times\Trans\rightarrow \natN$.
\end{definition}

\begin{definition}[distincT-PPNs]
    A distinctT-PPN is a PPN in which the set of parameters used in the $\Pre$ function and the set of those used in the $\Post$ function are disjoint.
\end{definition}

We could also consider classic Petri nets in which the initial marking is parameterised. The corresponding formalism is called P-PPN. Such a parametric initial marking is easily simulated with an initial transition that has parametric weights in its post arcs and sets the initial marking. Interestingly, it can also be proved that postT-PPNs can be (weakly) simulated by P-PPNs~\cite{DJLR15}.

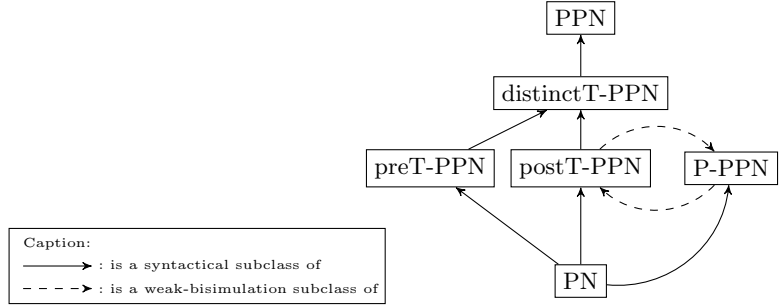
\begin{figure} \centering

\begin{tikzpicture}[node distance=1.3cm,>=stealth',bend angle=45,auto]

  \begin{scope}

	\node[draw] (T-PPN) at (1,-1) {PPN};
	\node[draw] (P-PPN) at (3,-3) {P-PPN};
	\node[draw] (distinctT-PPN) at (1,-2) {distinctT-PPN};
	\node[draw] (preT-PPN) at (-1,-3) {preT-PPN};
	\node[draw] (postT-PPN) at (1,-3) {postT-PPN};
      \node[draw] (PN) at (1,-4.5) {PN};
 
	\draw[->,draw=black] (PN) to (preT-PPN); 
	\draw[->,draw=black] (PN) to (postT-PPN); 
      \draw[->,draw=black] (PN) edge[bend right] (P-PPN); 
	
	\draw[->,draw=black] (postT-PPN) to (distinctT-PPN); 
	\draw[->,draw=black] (preT-PPN) to (distinctT-PPN); 

	\draw[->,draw=black] (distinctT-PPN) to (T-PPN); 
	
      \draw[->,dashed] (P-PPN) edge[bend left] (postT-PPN);
      \draw[->,dashed] (postT-PPN) edge[bend left] (P-PPN);
  \end{scope}

  \begin{scope}[xshift=-3.5cm,yshift=-1.8cm]
  \draw  (-3.1,-2) rectangle (1.9,-3); 
  \tiny
	\node (capt) at (-3,-2.2) [right] {Caption:};
	\node (c1) at (-3,-2.5) {};
	\node (c2) at (-3,-2.8) {};
	\node (capt1) at (-2,-2.5) [right] {: is a syntactical subclass of};
	\node (capt2) at (-2,-2.8) [right]{: is a weak-bisimulation subclass of};
	\draw[->] (c1) to (capt1);  
	\draw[->,dashed] (c2) to (capt2);  

\end{scope}
\end{tikzpicture}
		\caption{Subclasses of PPNs}
		\label{fig:subClasses}
	\end{figure}

\cref{fig:subClasses}~\cite{DJLR15} summarises this hierarchy of subclasses.

\subsection{Global Results}

We now summarise the current state-of-the art for the study of PPNs. \cref{tab:recapConcl1} gives the decidability results and whenever relevant the complexities for the universal problems, while \cref{tab:recapConcl2} gives them for the existential problems.
   \begin{table}[!!htb]
	\begin{center}
	\hspace*{-1cm}
	{\renewcommand{\arraystretch}{1.5}
	\scalebox{0.95}{
	\begin{tabular}{|c |c | c | c|}%
		\hline
            \cellHeader{Class} & \cellHeader{Reachability} & \cellHeader{S. Unboundness} & \cellHeader{Coverability} 
			\\	
		\hline
			PPN 
             &\cellUndec{}Undecidable  \tiny{\cite{DJLR15}} &\cellUndec{}Undecidable  \tiny{\cite{DJLR15}} & \cellUndec{}Undecidable \tiny{\cite{DJLR15}} 
			\\
		\hline
			preT-PPN 
             &\cellUndec{}Undecidable \tiny{\cite{David17}} &\cellDec{}\EXPSPACE-c \tiny{\cite{DJLR17}} &\cellDec{}\EXPSPACE-c \tiny{\cite{DJLR17}} 
			\\
		\hline
			postT-PPN 
			 &\cellUndec{}Undecidable \tiny{\cite{David17}} &\cellDec{}\EXPSPACE-c\tiny{\cite{David17}} &\cellDec{}\EXPSPACE-c \tiny{\cite{DJLR15}} 
			\\
		\hline
			P-PPN 
            & \cellOpen{} &\cellDec{}\EXPSPACE-c\tiny{\cite{David17}} &\cellDec{}\EXPSPACE-c \tiny{\cite{DJLR15}}
			\\
		\hline
			distinctT-PPN 
			 &\cellUndec{}Undecidable \tiny{\cite{David17}} &\cellDec{}\EXPSPACE-c\tiny{\cite{David17}} &\cellDec{}\EXPSPACE-c \tiny{\cite{DJLR15}}
			\\			
		\hline
	\end{tabular}
	}}
	\end{center}
	\caption{Un(decidability) and complexity results for the universal parametric problems}
	\label{tab:recapConcl1}
\end{table}

\begin{table}[!!htb]
	\begin{center}
	\hspace*{-1cm}
	{\renewcommand{\arraystretch}{1.5}
	\scalebox{0.95}{
        \begin{tabular}{|c |c | c | c |}%
		\hline
            \cellHeader{Class}& \cellHeader{Reachability} & \cellHeader{S. Unboundedness} & \cellHeader{Coverability} 
			\\	
		\hline
			PPN 
			 & \cellUndec{}Undecidable \tiny{\cite{DJLR15}} & \cellUndec{}Undecidable \tiny{\cite{DJLR15}}& \cellUndec{}Undecidable \tiny{\cite{DJLR15}}
			\\
		\hline
			preT-PPN 
             & \cellUndec{}Undecidable \tiny{\cite{David17}}& \cellDec{}\EXPSPACE-c\tiny{\cite{David17}} &\cellDec{}\EXPSPACE-c \tiny{\cite{DJLR15}}
			\\
		\hline
			postT-PPN 
            &\cellUndec{}Undecidable \tiny{\cite{David17}} & \cellDec{}\EXPSPACE-h\tiny{\cite{David17}}  &\cellDec{}\EXPSPACE-c \tiny{\cite{DJLR15}} 
			\\
		\hline
			P-PPN 
            & \cellOpenB{}Decidable \tiny{\cite{DJLR15}}& \cellDec{}\EXPSPACE-h\tiny{\cite{David17}} &\cellDec{}\EXPSPACE-c \tiny{\cite{DJLR15}}
			\\
		\hline
			distinctT-PPN 
             &\cellUndec{}Undecidable  \tiny{\cite{David17}} & \cellDec{}\EXPSPACE-h\tiny{\cite{David17}} &\cellDec{}\EXPSPACE-c \tiny{\cite{DJLR17}} 
			\\			
		\hline
	\end{tabular}
	}}
	\end{center}
	\caption{Un(decidability) and complexity results for the existential parametric problems}
	\label{tab:recapConcl2}
\end{table}

In order to establish the decidability results, we use a variety of techniques.  The most basic is that preT- and postT-PPNs have a strong monotonicity property~\cite{DJLR15}: increasing (resp. decreasing) the  value of parameters in a postT-PPN (resp. preT-PPN) can only add behaviours. Second we can reduce universal coverability in preT-PPNs to simultaneous unboundedness, and existential coverability in postT-PPNs to coverability in the $\omega$ Petri nets of~\cite{GHPR13}. Both of these reductions can be found in~\cite{DJLR17}. Finally, we can adapt the Karp \& Miller algorithm~\cite{KM69} to preT- and postT-PPNs~\cite{David17}.

\cref{tab:recapConcl3} presents results for the synthesis problem~\cite{DJLR17}. In
the cases where we can compute the set of adequate valuations (\cmark), we mostly rely on the use of an algorithm to compute upward-closed sets by Valk and Jantzen~\cite{VJ85}. The negative results (\xmark) come from the fact that the emptiness or the universality of the set cannot be decided as a direct consequence of the undecidability results above, so there is little hope to find a useful representation of that set. The case of distinctT-PPNs is similar in so far as if we can compute the solution and test its intersection with equality constraints we can solve the synthesis problem for any PPN, by replacing parameters used both in pre and post arcs by different parameters (which gives a distinctT-PPN) and then constraining the solution set with equality constraints on these different parameters.

   \begin{table}[!!htb]
	\begin{center}
	\hspace*{-1cm}
	{\renewcommand{\arraystretch}{1.5}
	\begin{tabular}{|c |c |c |c |}
		\hline
            \cellHeader{Class} & \cellHeader{Reachability} & \cellHeader{S. Unboundedness} & \cellHeader{Coverability} \\	
		\hline
			PPN 
             & \cellUndec{}\xmark & \cellUndec{}\xmark & \cellUndec{}\xmark
			\\
		\hline
			preT-PPN 
             & \cellUndec{}\xmark & \cellDec{}\cmark & \cellDec{}\cmark
			\\
		\hline
			postT-PPN 
            & \cellUndec{}\xmark & \cellDec{}\cmark& \cellDec{}\cmark 
			\\
		\hline
			P-PPN 
             & \cellOpen{} & \cellDec{}\cmark& \cellDec{}\cmark
			\\
		\hline
			distinctT-PPN 
             & \cellUndec{}\xmark & \cellUndec{}\xmark & \cellUndec{}\xmark
			\\			
		\hline
	\end{tabular}
	}
	\end{center}
       \caption{Results for the synthesis problem}
	\label{tab:recapConcl3}
	\end{table}
	
\subsection{Conclusion}
Parametric Petri nets are a powerful formalism to model flexible systems. In the general case, the interesting problems are undecidable but still useful subclasses can be obtained by restricting the use of parameters.
For most of these subclasses, it is possible to actually synthesise the values of
the parameters such that the net is unbounded, or such that some marking is coverable.
It would nevertheless be interesting to design semi-algorithms or incomplete algorithms
for the most expressive cases that do not fit in this restricted setting. A problem
also remains open, \ie{} the decidability of universal reachability for Petri nets
with a parameterised initial marking.

\section{Action synthesis}
\label{sec:action}

One of the classical approaches to verification and specification
of concurrent systems employs Kripke structures as models and
branching-time logics such as CTL as property description languages.
In contrast to the models presented earlier, Kripke structures allow
only for specifying sequential behaviours. 
Advanced data structures such as
Binary Decision Diagrams~\cite{Bryant86} (BDDs) together with algorithms
based on fixed-point specification of CTL enable efficient
verification of models whose state spaces exceed $10^{20}$~\cite{BurchCMDH90}. 
Here, we extend Action-Restricted Computation Tree
Logic~\ARCTL~\cite{Pecheur06} and its models with parameters,
to obtain a framework that benefits from BDD-based fixed-point
algorithms.

\subsection{Mixed Transition Systems}
Mixed Transition Systems~\cite{Pecheur06} (\mts) are essentially Kripke structures
with transitions labelled by actions. 

\begin{definition}[\mts]\label{mtsdef}
Let $\PV$ be a set of propositional variables. 
A {\em Mixed Transition System} is a 5-tuple 
$\model = (\modelSTS, \modelStInit, \modelACTS, \modelTr, \modelVS)$, where:
\begin{itemize}
\item $\modelSTS$ is a finite set of states, and
$\modelStInit \in \modelSTS$ is the initial state,
\item $\modelACTS$ is a non-empty finite set of actions,
\item $\modelTr \subseteq \modelSTS \times \modelACTS \times \modelSTS$
  is a transition relation,
\item $\modelVS : \modelSTS \to 2^\PV$ is a (state) valuation function.
\end{itemize}
\end{definition}
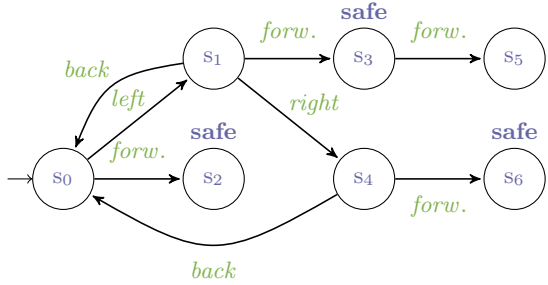
\begin{figure}[t!]
\centering
\begin{tikzpicture}[scale = 0.8]
\tikzstyle{every state} = []

\node[state,initial,label={[label distance=1pt]120:{}}] (s0) at (-1.0, -1.0) {$\styleloc{s_0}$};
\node[state,label=90:{}] (s1) at (1.5, 1.0) {$\styleloc{s_1}$};
\node[state,label=90:{\styleprop{\safe}}] (s2) at (1.5, -1.0) {$\styleloc{s_2}$};
\node[state,label=90:{\styleprop{\safe}}] (s3) at (4.0, 1.0) {$\styleloc{s_3}$};
\node[state,label=90:{}] (s4) at (4.0, -1.0) {$\styleloc{s_4}$};
\node[state,label=90:{}] (s5) at (6.5, 1.0) {$\styleloc{s_5}$};
\node[state,label=90:{\styleprop{\safe}}] (s6) at (6.5, -1.0) {$\styleloc{s_6}$};

\path [->,>=stealth',shorten >=1pt, auto, node distance=7cm, semithick]
(s0) edge node[above=0.1, pos=0.4] {\styleact{\leftp}} (s1)
(s1) edge[above, looseness=1.3, bend right, pos=0.5, text width=1.3cm] node {\styleact{\backp}} (s0)
(s0) edge node[above=0.1, pos=0.5] {\styleact{\forwardp}} (s2)
(s1) edge node[above=0.1, pos=0.5] {\styleact{\forwardp}} (s3)
(s3) edge node[above=0.1, pos=0.5] {\styleact{\forwardp}} (s5)
(s4) edge node[below=0.1, pos=0.5] {\styleact{\forwardp}} (s6)
(s1) edge node[above=0.2, pos=0.75] {\styleact{\rightp}} (s4)
(s4) edge[below, bend left, looseness=1.4] node [below=0.1, pos=0.5] {\styleact{\backp}} (s0)
;

\end{tikzpicture}
{
\caption[Simple mixed transition system]{A simple mixed transition system}\label{runnexfig}.
}
\end{figure}

We write $s\stackrel{a}{\to}s'$ if $(s,a,s')\in \modelTr$. 
Let $\concractset \subseteq \modelACTS$ %
and $\pi = (s_0,a_0,s_1,a_1,\ldots)$ be a finite or infinite sequence of 
interleaved states and actions. By $|\pi|$ we denote the number of 
states in $\pi$ if $\pi$ is finite, and $\omega$ if $\pi$ is infinite. 
A sequence $\pi$ is a \emph{path} over $\concractset$ iff 
$s_i \stackrel{a_i}{\to} s_{i+1}$ and $a_i \in \concractset$ for each $i < |\pi|$
and either $\pi$ is infinite or its final state does not have a $\concractset$-successor 
state in $\modelSTS$, \ie{} $\pi = (s_0,a_0,s_1,a_1,\ldots, s_m)$ for some $m\in\grandn$
and there is 
no $s'\in\modelSTS$ and $a\in\concractset$ such that $s_m\stackrel{a}{\to}s'$.
By $\pi_i$ we denote the $i$--th state of $\pi$ for all $i\in\grandn$.

The set of all paths over $\concractset$ 
in $\model$ is denoted by $\modelPaths(\model, \concractset)$,
and the set of all paths $\pi\in\modelPaths(\model, \concractset)$ 
starting from a given state $s\in\modelSTS$ is denoted by 
$\modelPaths(\model, \concractset, s)$.
We typically omit the symbol $\model$,
writing $\modelPaths(\concractset)$
and $\modelPaths(\concractset,s)$.
By $\modelPaths^\omega(\concractset)$ and $\modelPaths^\omega(\concractset,s)$
we mean the corresponding sets restricted to the infinite paths~only.

\begin{example}[\mts]\label{runnex1}
A \mts\ with $\PV = \{\styleprop{p},\styleprop{\safe}\}$,
$\modelACTS = \{\styleact{\leftp}, \styleact{\rightp}, \styleact{\forwardp}, \styleact{\backp}\}$,
and initial state $\styleloc{s_0}$ is shown in \cref{runnexfig}.
The path $\pi = (\styleloc{s_0},\styleact{\leftp},\styleloc{s_1},\styleact{\rightp},\styleloc{s_4})$ belongs to $\modelPaths(\{\styleact{\leftp}, \styleact{\rightp}\})$, 
but not to the set $\modelPaths(\{\styleact{\leftp}, \styleact{\rightp}, \styleact{\backp}\})$.
The reason is that while $\pi$ is a maximal path over $\{\styleact{\leftp}, \styleact{\rightp}\}$,
it is not maximal over $\{\styleact{\leftp}, \styleact{\rightp}, \styleact{\backp}\}$ as it can be extended e.g. into
the infinite path 
$\pi' = (\styleloc{s_0},\styleact{\leftp},\styleloc{s_1},\styleact{\rightp},\styleloc{s_4},\styleact{\backp},\styleloc{s_0}, 
\styleact{\leftp},\styleloc{s_1},\styleact{\rightp},\styleloc{s_4},\styleact{\backp},\styleloc{s_0}, \ldots)
\in\modelPaths(\{\styleact{\leftp}, \styleact{\rightp}, \styleact{\backp}\})$.
\end{example}

\subsection{Parametric Action-Restricted CTL}

The main difference 
between \ARCTL\ and CTL is that in \ARCTL\ each path quantifier is subscripted with
a set of actions, \eg{}
$
E_{\{\styleact{\leftp}, \styleact{\rightp}\}}G (E_{\{\styleact{\forwardp}\}}F \styleprop{\safe})
$
may be read as \emph{``there exists a path over \styleact{\leftp}\ and \styleact{\rightp}, on which it 
holds globally that a state satisfying \styleprop{\safe}\ is reachable along some path
over \styleact{\forwardp}''.}

Parametric \ARCTL\ (\PARCTL) extends \ARCTL\ by allowing free variables in place 
of sets of actions, \eg{} $E_YG (E_ZF\styleprop{\safe})$ is a formula of \PARCTL,
where $Y$ and $Z$ are free variables.

\begin{definition}[\PARCTL\ syntax]
Let $\modelACTS$ be a finite set of actions, 
$\ActVariables$ a finite set of variables, 
$\PV$ a set of propositional variables, 
and $\ActSets = 2^\modelACTS \setminus \{\emptyset\}$.
The set of formulae of \PARCTL\ is defined by the following grammar: 
\[\phi := p \mrule \neg\phi \mrule \phi\lor\phi \mrule 
E_\alpha X\phi \mrule E_\alpha G\phi \mrule E^{\omega}_\alpha\Gom \phi \mrule
E_\alpha(\phi\;U \phi),\]
where $p\in\PV$ and $\alpha\in\concsets$.
\end{definition}
The basic path quantifiers and modalities of \PARCTL\ have the same 
meaning as in CTL. 
The superscript $^\omega$ restricts the quantification to 
the infinite paths, whereas the subscript $_\alpha$ restricts the quantification to the paths over $\alpha$.

The semantics of \PARCTL\ is defined \wrt{} parameter valuations,
\ie{} functions $\upsilon\colon \ActVariables\to \ActSets$.  $\ParVals$
denotes the set of all parameter valuations.
For conciseness, for $\upsilon\in\ParVals$ we write
$\upsilon(\alpha) = \concractset$ if $\alpha = \concractset\subseteq\modelACTS$
and
$\upsilon(\alpha) = \upsilon(Y)$ if $\alpha = Y\in\ActVariables$.
Moreover, we assume that $O^\epsilon = O$, for $O\in\{E,A,\modelPaths\}$.

\begin{definition}[\PARCTL\ semantics]\label{parctlsemdef}
Let $\model = (\modelSTS, \modelStInit, \modelACTS, \modelTr, \modelVS)$ be an \mts\
and $\upsilon\in\ActVals$ a parameter valuation. 
The relation $\models_\upsilon$ is defined as follows:
\begin{itemize}
\item $s\models_\upsilon p$ iff $p\in\modelVS(s)$,
\item $s\models_\upsilon \neg\phi$ iff $s\not\models_\upsilon \phi$,
\item $s\models_\upsilon \phi\lor\psi$ iff $s\models_\upsilon \phi$ or $s\models_\upsilon \psi$,

\item $s\models_\upsilon E_\alpha X\phi$ iff $|\pi|> 1$ and $\pi_1\models_\upsilon \phi$, 
      for some $\pi\in\modelPaths(\upsilon(\alpha), s)$,

\item $s\models_\upsilon E^r_\alpha G\phi$ iff $\pi_i\models_\upsilon \phi$ 
for all $i < |\pi|$, for some $\pi\in\modelPaths^r(\upsilon(\alpha), s)$,

\item $s\models_\upsilon E_\alpha(\phi\;U\psi)$ iff $\pi_i\models_\upsilon\psi$ 
      for some  $i < |\pi|$ and $\pi_j\models_\upsilon\phi$ for all $0 \le j <i$, 
			for some $\pi\in\modelPaths(\upsilon(\alpha),s)$,

\end{itemize}
where $p\in\PV$, $\phi,\psi\in\PARCTL$, $r\in\{\omega, \epsilon\}$,
and $\alpha\in\concsets$.
\end{definition}

\begin{example}
For the \mts\ in \cref{runnexfig} we have $\styleloc{s_0}\models_\upsilon E_YG (E_ZF
\styleprop{\safe})$
iff $\styleact{\forwardp}\in\upsilon(Z)$.
\end{example}

\subsection{Parameter Synthesis for \PARCTL}

In this subsection we show how to recursively characterise \PARCTL\
using the basic operator of parametric pre-image. These equivalences give
rise to fixed-point algorithms that can be implemented using BDDs.

Let $\phi\in\PARCTL$. Our goal is to construct 
the function $f_\phi\colon\modelSTS\to 2^{\ParVals}$ s.t.
for all $s\in\modelSTS$ we have
$
s\models_\upsilon\phi \text{ iff } \upsilon\in f_\phi(s)
$.
In other words, the set $f_\phi(s)$ consists of all the parameter valuations that make $\phi$ true in state $s$.
Let us show how to build this function recursively, case by case.
We omit the treatment of non-parametric modalities
as the classical non-parametric methods of symbolic verification
carry here with minimal alterations~\cite{HuthRyan-logic,RaimondiL07}.

\paragraph{Boolean Connectives and Non-parametric Modalities.} 

For each $p\in\PV$, $\phi,\psi\in\PARCTL$ we have
$f_p(s) = \ActVals$  if $p\in\modelVS(s)$ and 
$f_p(s) = \emptyset$ otherwise;
$f_{\neg\phi}(s) = \ActVals\setminus f_\phi(s)$;
and $f_{\phi\lor\psi}(s) = f_\phi(s) \cup f_\psi(s)$.

\paragraph{Parametric Pre-image and NeXt.}
Let $f\colon\modelSTS\to 2^\ActVals$ be a function.
The \emph{parametric pre-image} of $f$ \wrt{} $Y\in\ActVariables$ is defined 
as the function $\preime_Y(f):\modelSTS\to 2^\ActVals$ s.t. 
$
\preime_Y(f)(s) = \left\{ 
\upsilon \mrule \exists_{s'\in\modelSTS}\;\exists_{a\in\upsilon(Y)} \;s\stackrel{a}{\to}s' \land \upsilon\in f(s')
\right\}
$ for each $s\in\modelSTS$.
Intuitively, for each $\phi \in \PARCTL$ in $\preime_Y(f_\phi)(s)$ we collect
all the parameter valuations $\upsilon$ s.t. some state $s'$ satisfying $s'\models_\upsilon\phi$ %
can be reached by firing
in $s\in\modelSTS$ an action from $\upsilon(Y)$.
We therefore have $f_{E_Y X\phi} = \preime_Y(f_\phi)(s)$

\paragraph{Parametric Temporal Modalities.}
We employ the following equations to
deal with two versions of the Globally modality:
\begin{align*}
& f_{E^\omega_Y\Gom\phi}(s) 
= f_{\phi}(s) \cap \preime_Y(f_{ E^\omega_Y\Gom\phi})(s),\\
& f_{E_YG\phi}(s) 
= f_{\phi}(s) \cap (\preime_Y(f_{ E^\omega_Y G\phi})(s) \cup f_{\neg E_YX\BTrue}(s)).
\end{align*}
The following equation characterises the Until modality:
\begin{align*}
f_{E_Y(\phi U\psi)}(s)  = f_{\psi}(s) \cup (f_{\phi}(s) \cap \preime_Y(f_{E_Y(\phi U\psi)})(s)).
\end{align*}

We refer to~\cite{KnapikMP15,KnapikP14} on how to turn the
above equations into fixed-point algorithms and implement them using BDDs.
In all the cases, the returned result of running the overall synthesis
algorithm for $\phi\in\PARCTL$ is the BDD that represents $f_\phi$.
This structure can be then queried for individual parameter valuations;
for a certain class of formulae it is possible to synthesise minimal
parameter valuations using prime implicants.

As a closing note let us mention that the emptiness problem for
$\PARCTL$, \ie{} the question whether $f_\phi(\modelStInit) \ne \emptyset$
for $\phi\in\PARCTL$ is known to be NP-complete~\cite{KnapikMP15}.

\section{Tools}
\label{sec:tools}

In this section, we finally briefly review tools related to the aforementioned formalisms.

\subsection{\imitator{}}
\label{sec:tool:imitator}

\imitator{}~\cite{AFKS12} is a software tool for parametric verification and robustness analysis of PTAs augmented with integer variables and stopwatches.
Parameters can be used both in the model and in the properties.
Verification capabilities include reachability-synthesis, deadlock-freeness-synthesis~\cite{Andre16}, non-Zeno model checking~\cite{ANPS17}, minimal-time synthesis~\cite{ABPV19},
and trace-preservation-synthesis.
\imitator{} is fully written in OCaml, and makes use of the Parma Polyhedra Library~\cite{BHZ08}.
It also features distributed capabilities to run over a cluster.

\imitator{} comes with a benchmarks library available under an open source license~\cite{Andre18FTSCS}.

\imitator{} was successfully used in several application domains such as 
	parametric schedulability analysis of a prospective architecture for the flight control system of the next generation of spacecrafts designed at ASTRIUM Space Transportation~\cite{FLMS12},
	formal timing analysis of music scores~\cite{FJ13},
	verification of software product lines~\cite{LSBL17},
	monitoring logs from the automotive domain against parametric properties~\cite{AHW18},
	and was used to propose a solution to a challenge related to a distributed video processing system by Thales~\cite{SAL15}.

\subsubsection{Related tools}

The first tool to support modelling and verification using parametric timed automata was \hytech{}~\cite{HHW97}.
In fact, \hytech{} supports linear hybrid automata (including clocks, parameters, stopwatches and general continuous variables); 
it can compute the state space, and perform operations (such as intersection, convex hull, difference) between sets of symbolic states.
Therefore, it can be used to perform parametric model checking using reachability checking~\cite{ABBL98}.
\hytech{} is not maintained anymore, but can still be found online in the form of a standalone binary for Linux.\footnote{%
	\url{https://embedded.eecs.berkeley.edu/research/hytech/}
}

In~\cite{HRSV02}, an extension of \uppaal{} implementing parametric difference bound matrices (PDBMs) and hence allowing for verification 
using PTAs is mentioned.
However, this tool does not seem to be available anywhere online.

PHAVer~\cite{Frehse2008} is a tool for verifying safety properties of hybrid systems.
It notably relies on exact arithmetic (with unlimited precision) using the Parma Polyhedra Library~\cite{BHZ08}; on the other hand, 
it also supports approximations.

SpaceEx~\cite{FLDCRLRGDM11} can be seen as a successor of PHAVer, and also tackles verification of reachability properties for hybrid systems.
Parameters are not natively supported, but can be encoded using variables that are arbitrarily set up upon system start, and then remain 
subsequently constant (with a 0-slope).
SpaceEx seems to have a lot of interesting recent developments.

\PSyHCoS{}~\cite{ALSDL13} allows the synthesis of parameters for a parametric extension of the process algebra Stateful timed CSP~\cite{SLDLSA13}, 
itself a timed extension of Hoare's communicating sequential processes~\cite{Hoare85}.
When compared to other formalisms such as (parametric) timed automata, (parametric) stateful timed CSP has the advantage of giving the designer 
the ability to specify hierarchical systems.

Finally, \symrob{} is not strictly a tool for synthesis, but allows robustness measurement for timed automata~\cite{Sankur15}.

\subsection{\romeo}
\label{sec:tool:romeo}

\romeo{}~\cite{LRST09} is a model-checking tool for a selection of hybrid extensions
of Petri nets, enriched with discrete variables. 
In particular, it supports parametric time Petri nets, a formalism shown to be close to PTAs in terms of expressiveness~\cite{BCHLR05,TLR09}.
\romeo{} allows the use of parametric linear expressions in the time intervals of the transitions, and the addition of linear constraints on the parameters to restrict their domain. 
\romeo{} provides a simulator and an integrated model-checker supporting a subset of parametric TCTL (including reachability-synthesis and unavoidability-synthesis), in which ``Until'' modalities cannot be nested. It also features optimal cost reachability and parameter synthesis for cost-bounded reachability.
\romeo{} implements in particular an original algorithm for integer parameter synthesis using a symbolic (continuous) representation~\cite{JLR15}.
\romeo{} is mainly written in C++, and makes use of the Parma Polyhedra Library~\cite{BHZ08}.
It has been successfully used in a few and diverse case-studies including the analysis of resilience properties in oscillatory biological systems~\cite{magnin-biosystems-16}; the synthesis of environment requirements for an aerial video tracking system~\cite{parquier-FTSCS-16}; and the analysis of operational scenarios modelling in the DGA OMOTESC project~\cite{seidner-phd-09}.

\subsection{Spatula}
\label{sec:tool:spatula}

Spatula~\cite{SPATULApage,KnapikMP15} implements the theory of action synthesis outlined in~\cref{sec:action}.
The tool is written in C++ and employs CUDD library~\cite{CUDDpage} for representing
and manipulating BDDs. 
Spatula accepts models represented as networks of automata written in a
simplified C-like input language and \PARCTL\ as property description language.
The result of synthesis for a given property is a BDD that represents all the valuations that 
make the property true. 
Spatula can also list all the minimal valuations (\wrt{} bitwise comparison)
for the existential part of the logic.

	\newcommand{\CCIS}{Communications in Computer and Information Science}
	\newcommand{\ENTCS}{Electronic Notes in Theoretical Computer Science}
	\newcommand{\FAC}{Formal Aspects of Computing}
	\newcommand{\FI}{Fundamenta Informaticae}
	\newcommand{\FMSD}{Formal Methods in System Design}
	\newcommand{\IJFCS}{International Journal of Foundations of Computer Science}
	\newcommand{\IJSSE}{International Journal of Secure Software Engineering}
	\newcommand{\IPL}{Information Processing Letters}
	\newcommand{\JLAP}{Journal of Logic and Algebraic Programming}
	\newcommand{\JLAMP}{Journal of Logical and Algebraic Methods in Programming} %
	\newcommand{\JLC}{Journal of Logic and Computation}
	\newcommand{\LMCS}{Logical Methods in Computer Science}
	\newcommand{\LNCS}{Lecture Notes in Computer Science}
	\newcommand{\RESS}{Reliability Engineering \& System Safety}
	\newcommand{\STTT}{International Journal on Software Tools for Technology Transfer}
	\newcommand{\TCS}{Theoretical Computer Science}
	\newcommand{\ToPNoC}{Transactions on Petri Nets and Other Models of Concurrency}
	\newcommand{\TSE}{{IEEE} Transactions on Software Engineering}

\ifdefined\AuthorVersion

	\renewcommand*{\bibfont}{\small}
	\printbibliography[title={References}]
\else
	\bibliographystyle{splncs04}
	\bibliography{main}

\fi

\end{document}